\documentclass[preprint,12pt]{elsarticle}

\input{Qcircuit}
\usepackage{appendix,amsmath,amssymb,graphicx,mathptmx,nccmath,subfigure}

\newcounter{MathematicaIO}
\newlength{\fxargpush}
\addtocounter{MathematicaIO}{1}

\begin{document}

\begin{frontmatter}

\title{An efficient quantum circuit analyser \\on qubits and qudits}
\author{T. Loke and J. B. Wang}
\address{School of Physics, The University of Western Australia, 6009 Perth, Australia}

\begin{abstract}

This paper presents a highly efficient decomposition scheme and its associated \emph{Mathematica} notebook for the analysis of complicated quantum circuits comprised of single/multiple qubit and qudit quantum gates.  In particular, this scheme reduces the evaluation of multiple unitary gate operations with many conditionals to just two matrix additions, regardless of the number of conditionals or gate dimensions.  This improves significantly the capability of a quantum circuit analyser implemented in a classical computer. This is also the first efficient quantum circuit analyser to include qudit quantum logic gates. 


\end{abstract}

\end{frontmatter}

\section{Program Summary}

\begin{flushleft}
{\sl Title of program}: CUGates.m \\

{\sl Programming language used}: \emph{Mathematica} \\
 
\hangindent=10pt {\sl Computers and operating systems}: any computer installed with \emph{Mathematica} 6.0 or higher\\

{\sl Distribution format}: \emph{Mathematica} notebook \\

\hangindent=10pt {\sl Nature of problem}: The \emph{CUGates} notebook simulates arbitrarily complex quantum circuits comprised of single/multiple qubit and qudit quantum gates.\\
 
\hangindent=10pt {\sl Method of solution}:  
It utilizes an irreducible form of matrix decomposition for a general controlled gate with multiple conditionals and is highly efficient in simulating complex quantum circuits.\\ 

\hangindent=10pt {\sl Running time}:  
Details of CPU time usage for various example runs are given in Section 4.\\ 

\hangindent=10pt {\sl Program obtainable from}: CPC Program Library, Queen�s University of Belfast, N. Ireland \\
\end{flushleft}

\section{Introduction}

At the heart of a quantum computer lies a set of qubits and/or qudits whose states are manipulated by a series of quantum logic gates, namely a quantum circuit, to provide the ultimate computational results.  It is therefore of particular interest to be able to efficiently evaluate the performance of a quantum circuit (such as its reliability, effectiveness, robustness, sensitivity to decoherence and errors) in the design stage using a classical computer. 

There are currently several quantum computer simulators reported in the literature \cite{Diaz2006,Radtke2005,Obenland1998,Raedt2007,Gutierrez2010}, which simulate quantum circuits consisting of 1, 2 or 3 qubit gates such as the Hadamard, CNOT and Toffoli gate. The CNOT and Toffoli gate are examples of controlled unitary gates (CUGs), which implement operations that are conditional on the state of the specified control qubits. Other more general CUGs (acting across qubits or qudits) can always be decomposed in terms of a universal set of 1- and 2-qubit quantum gates \cite{Nielsen2000}, but this would require significant computational overhead in the analysis. To the best of our knowledge, there are no efficient quantum simulators on quantum circuits with multiple qudit controlled quantum gates.

In this paper, we present a highly efficient scheme for the evaluation of arbitrary CUGs. This scheme reduces the evaluation of multiple unitary gate operations with many conditionals to just two matrix additions, regardless of the number of conditionals or gate dimensions. The implementation of this scheme, and many other functions used to analyse quantum circuits, is provided in a \emph{Mathematica} 7.0 package entitled \emph{CUGates.m}. The computation time required to evaluate the CNOT and Toffoli gates using this package is compared with the \emph{QDENSITY} package \cite{Diaz2006} and is found to be several orders of magnitude more efficient. Examples of quantum circuits involving controlled unitary gates and their analysis using the notebook are presented. A compilation of the \emph{Mathematica} code presented in the paper is provided in the \emph{Mathematica} notebook \emph{CUGates.nb}.

\section{Decomposition of CUGs}

\subsection{CUGs across qubits}

\subsubsection{Definitions and notation}

Denote a set of qubits as $ Q $, and the wavefunction (if definable) for the $ i $th qubit as $ \ket{\psi_i} $. $ Q $ is in a basis state iff $ \ket{\psi_i} = \ket{0} \vee \ket{\psi_i} = \ket{1} $ $ \forall $ $ i \in Q $. Define $ C^{c_i} $ as being conditional on the state $ \ket{1} $ of qubit $ c_i $, and $ \bar{C}^{\bar{c}_j} $ as being conditional on the state $ \ket{0} $ of qubit $ \bar{c}_j $.

\begin{figure}[h]
\centering
\mbox
{
\Qcircuit @C=1.00em @R=1.50em 
{
	\lstick{\ket{\psi_1}} & \ctrl{1}           & \qw \\
	\lstick{\ket{\psi_2}} & \gate{U_1}         & \qw \\
	\lstick{\ket{\psi_3}} & \ctrlo{1} \qwx     & \qw \\
	\lstick{\ket{\psi_4}} & \multigate{1}{U_2} & \qw \\
	\lstick{\ket{\psi_5}} & \ghost{U_2}        & \qw \\
	\lstick{\ket{\psi_6}} & \ctrlo{-1}         & \qw
}
}
\caption{The $ C^{1}\bar{C}^{3,6}U^{2}_{1}U^{4}_{2} $ gate, with $ C $ conditional on qubit 1 being $ \ket{1} $, $ \bar{C} $ conditional on qubit 3 and 6 being $ \ket{0} $, and the operations $ U_{1} $ and $ U_{2} $ are implemented on qubits 2 and 4 to 5 respectively.}
\label{fig:CUGQubitEg}
\end{figure}
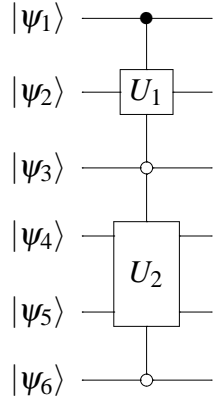

A CUG with conditionals $ C^{c_1,...,c_i}\bar{C}^{\bar{c}_1,...,\bar{c}_j} $  implementing unitary operations $ U^{u_1}_{1}...U^{u_k}_{k} $, where $ u_1,..., u_k $ denotes the starting qubit of the corresponding $ U $ block, is represented by $ C^{c_1,...,c_i}\bar{C}^{\bar{c}_1,...,\bar{c}_j}U^{u_1}_{1}...U^{u_k}_{k} $. Effectively, the action of this CUG is such that it implements the operations $ U^{u_1}_{1}...U^{u_k}_{k} $ iff the set of control qubits is in the basis state described by $ \ket{\psi_{c_1,...,c_i}} = \ket{1} $ and $ \ket{\psi_{\bar{c}_1,...,\bar{c}_j}} = \ket{0} $. For any other basis state, the CUG leaves the system of qubits unchanged. Figure \ref{fig:CUGQubitEg} shows an example of the $ C^{1}\bar{C}^{3,6}U^{2}_{1}U^{4}_{2} $ gate.

\subsubsection{Decomposition}

\begin{figure}[h]
\centering
\mbox
{
\Qcircuit @C=0.75em @R=2.25em 
{
	\lstick{\ket{\psi_1}} & \ctrl{1} & \qw \\
	\lstick{\ket{\psi_2}} & \gate{U} & \qw
}
}
\caption{The $ CU $ gate.}
\label{fig:CU}
\end{figure}
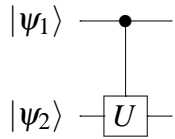

An efficient way to evaluate arbitrary controlled unitary gates is to decompose the operation by defining the projection operators $ P_0 $ and $ P_1 $ as:
\begin{equation}
\nonumber
P_0 = 
\left(
\begin{array}{ c c }
	1 & 0 \\
	0 & 0 \\
\end{array}
\right)
, \mbox{ } P_1 = 
\left(
\begin{array}{ c c }
	0 & 0 \\
	0 & 1 \\
\end{array}
\right).
\end{equation}
Note that $ P_0 $ and $ P_1 $ are non-unitary matrices and $ P_0 + P_1 = I_2 $ is the 2-by-2 identity matrix. Now consider the $ C^{1}U^{2} $ (abbreviated as $ CU $) gate, shown in Figure \ref{fig:CU}.

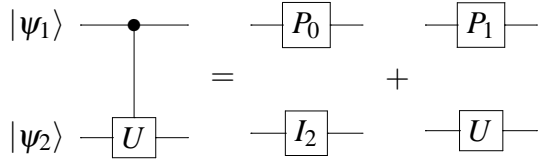
\begin{figure}[h]
\centering
\mbox
{
\Qcircuit @C=1.00em @R=2.80em
{
	\lstick{\ket{\psi_1}} & \ctrl{1} & \qw \\
	\lstick{\ket{\psi_2}} & \gate{U} & \qw 
}
\qquad
\Qcircuit @C=1.00em @R=1.00em
{
	           & \gate{P_0} & \qw \\
	\lstick{=} &            &     \\
	           & \gate{I_2} & \qw 
}
\qquad
\Qcircuit @C=1.00em @R=1.00em
{
	           & \gate{P_1} & \qw \\
	\lstick{+} &            &     \\
	           & \gate{U}   & \qw 
}
}
\caption{Decomposition of the $ CU $ gate.}
\label{fig:CUDecompose}
\end{figure}

It can be readily verified and proven that the matrix for the $ CU $ gate is given as $ CU = P_0 \otimes I_2 + P_1 \otimes U$  (see appendix A for details). This result, called the decomposition of the $ CU $ gate as a sum, is graphically summarised in Figure \ref{fig:CUDecompose}.

The key idea is that we can use the projection operators $ P_0 $ and $ P_1 $ to project the set of control qubits to a basis state. For any basis state, the action of the CUG gate is either just the $ U^{u_1}_{1}...U^{u_k}_{k} $ operators, or no action at all (i.e. the identity operator). By considering all possible basis states of the set of control qubits, we can construct the matrix of the CUG gate by summing together the action of the CUG gate corresponding to each possible basis state.

For any arbitrary $ C^{c_1,...,c_i}U^{u_1}_{1}...U^{u_k}_{k} $ gate, consider replacing each conditional with a $ P_0 $ or $ P_1 $ operator. This can be done in $ 2^i $ distinct ways. For the basis state described by $ \ket{\psi_{c_1,..., c_i}} = \ket{1} $, which corresponds to the permutation $ C^{m} \rightarrow P_1 $ $ \forall $ $ m = c_1,...,c_i $, the action of the CUG is the operations $ U^{u_1}_{1}...U^{u_k}_{k} $. Any other basis state (and hence permutation) corresponds to the action of the CUG being the identity operator. The sum of the $ 2^i $ permutations yields the matrix of the $ C^{c_1,..., c_i}U^{u_1}_{1}...U^{u_k}_{k} $ gate. For example, 
$$ C^{1,3}U^{2} = P_0 \otimes I_2 \otimes P_0 + P_0 \otimes I_2 \otimes P_1 + P_1 \otimes I_2 \otimes P_0 + P_1 \otimes U \otimes P_1 , $$
as graphically shown in Figure \ref{fig:C2UDecompose} (see appendix B for a mathematical proof).

\begin{figure}[h]
\centering
\mbox
{
\Qcircuit @C=1.00em @R=1.75em 
{
	\lstick{\ket{\psi_1}} & \ctrl{1}  & \qw \\
	\lstick{\ket{\psi_2}} & \gate{U}  & \qw \\
	\lstick{\ket{\psi_3}} & \ctrl{-1} & \qw
}
\qquad
\Qcircuit @C=1.00em @R=1.00em 
{
	           & \gate{P_0} & \qw \\
	\lstick{=} & \gate{I_2} & \qw \\
	           & \gate{P_0} & \qw
}
\qquad
\Qcircuit @C=1.00em @R=1.00em 
{
	           & \gate{P_0} & \qw \\
	\lstick{+} & \gate{I_2} & \qw \\
	           & \gate{P_1} & \qw
}
\qquad
\Qcircuit @C=1.00em @R=1.00em 
{
	           & \gate{P_1} & \qw \\
	\lstick{+} & \gate{I_2}  & \qw \\
	           & \gate{P_0} & \qw
}
\qquad
\Qcircuit @C=1.00em @R=1.00em 
{
	           & \gate{P_1} & \qw \\
	\lstick{+} & \gate{U}   & \qw \\
	           & \gate{P_1} & \qw
}
}
\caption{Decomposition of the $ C^{1,3}U^{2} $ gate.}
\label{fig:C2UDecompose}
\end{figure}
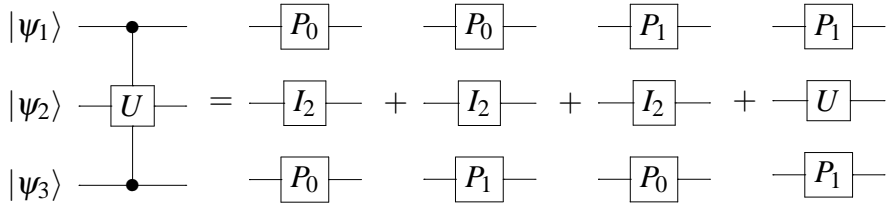

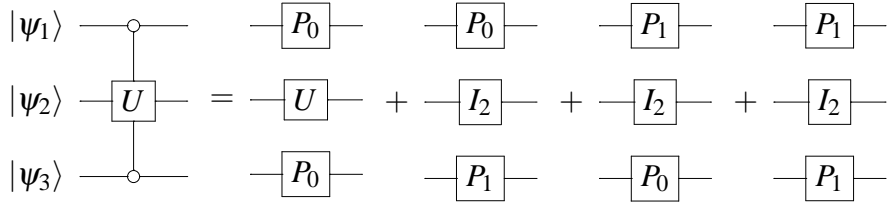
\begin{figure}[h]
\centering
\mbox
{
\Qcircuit @C=1.00em @R=1.55em 
{
	\lstick{\ket{\psi_1}} & \ctrlo{1}  & \qw \\
	\lstick{\ket{\psi_2}} & \gate{U}   & \qw \\
	\lstick{\ket{\psi_3}} & \ctrlo{-1} & \qw
}
\qquad
\Qcircuit @C=1.00em @R=1.00em 
{
	           & \gate{P_0} & \qw \\
	\lstick{=} & \gate{U}   & \qw \\
	           & \gate{P_0} & \qw
}
\qquad
\Qcircuit @C=1.00em @R=1.00em 
{
	           & \gate{P_0} & \qw \\
	\lstick{+} & \gate{I_2} & \qw \\
	           & \gate{P_1} & \qw
}
\qquad
\Qcircuit @C=1.00em @R=1.00em 
{
	           & \gate{P_1} & \qw \\
	\lstick{+} & \gate{I_2} & \qw \\
	           & \gate{P_0} & \qw
}
\qquad
\Qcircuit @C=1.00em @R=1.00em 
{
	           & \gate{P_1} & \qw \\
	\lstick{+} & \gate{I_2} & \qw \\
	           & \gate{P_1} & \qw
}
}
\caption{Decomposition of the $ \bar{C}^{1,3}U^{2} $ gate.}
\label{fig:Cbar2UDecompose}
\end{figure}

Similarly, for any arbitrary $ \bar{C}^{\bar{c}_1,...,\bar{c}_j}U^{u_1}_{1}...U^{u_k}_{k} $ gate, consider the $ 2^j $ possible permutations that arise from replacing each $ \bar{C} $ conditional with a $ P_0 $ or $ P_1 $ operator. For the basis state described by $ \ket{\psi_{\bar{c}_1,...,\bar{c}_i}} = \ket{0} $, which corresponds to the permutation $ \bar{C}^{n} \rightarrow P_0 $ $ \forall $ $ n = \bar{c}_1,...,\bar{c}_j $, the action of the CUG is the operations $ U^{u_1}_{1}...U^{u_k}_{k} $. Any other basis state corresponds to the action of the CUG being the identity operator. The sum of the $ 2^j $ permutations gives the matrix of the $ \bar{C}^{\bar{c}_1,...,\bar{c}_j}U^{u_1}_{1}...U^{u_k}_{k} $ gate, for example,
$$ \bar{C}^{1,3}U^{2}  = P_0 \otimes U \otimes P_0 + P_0 \otimes I_2 \otimes P_1 + P_1 \otimes I_2 \otimes P_0 + P_1 \otimes I_2 \otimes P_1 , $$
as graphically shown in Figure \ref{fig:Cbar2UDecompose}.

\begin{figure}[h]
\centering
\mbox
{
\Qcircuit @C=1.00em @R=1.60em 
{
	\lstick{\ket{\psi_1}} & \ctrlo{1} & \qw \\
	\lstick{\ket{\psi_2}} & \gate{U}  & \qw \\
	\lstick{\ket{\psi_3}} & \ctrl{-1} & \qw
}
\qquad
\Qcircuit @C=1.00em @R=1.00em 
{
	           & \gate{P_0} & \qw \\
	\lstick{=} & \gate{I_2} & \qw \\
	           & \gate{P_0} & \qw
}
\qquad
\Qcircuit @C=1.00em @R=1.00em 
{
	           & \gate{P_0} & \qw \\
	\lstick{+} & \gate{U}   & \qw \\
	           & \gate{P_1} & \qw
}
\qquad
\Qcircuit @C=1.00em @R=1.00em 
{
	           & \gate{P_1} & \qw \\
	\lstick{+} & \gate{I_2} & \qw \\
	           & \gate{P_0} & \qw
}
\qquad
\Qcircuit @C=1.00em @R=1.00em 
{
	           & \gate{P_1} & \qw \\
	\lstick{+} & \gate{I_2} & \qw \\
	           & \gate{P_1} & \qw
}
}
\caption{Decomposition of the $ C^{3}\bar{C}^{1}U^{2} $ gate.}
\label{fig:CCbarUDecompose}
\end{figure}
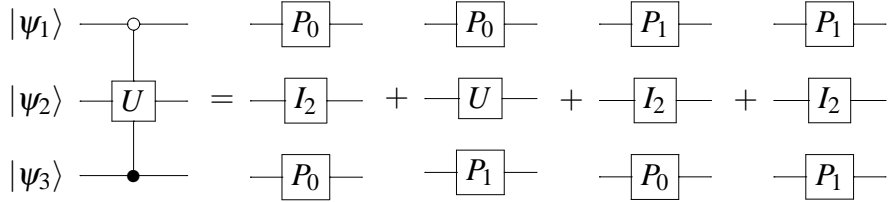

Hence, for any arbitrary $ C^{c_1,..., c_i}\bar{C}^{\bar{c}_1,...,\bar{c}_j}U^{u_1}_{1}...U^{u_k}_{k} $ gate, we consider each of the $ 2^{i+j} $ permutations that arise from replacing each $ C $ and $ \bar{C} $ conditional with a $ P_0 $ or $ P_1 $ operator. For the basis state described by $ \ket{\psi_{c_1,..., c_i}} = \ket{1} $ and $ \ket{\psi_{\bar{c}_1,...,\bar{c}_i}} = \ket{0} $, which corresponds to the permutation $ C^{m} \rightarrow P_1 $ $ \forall $ $ m = c_1,...,c_i $ and $ \bar{C}^{n} \rightarrow P_0 $ $ \forall $ $ n = \bar{c}_1,...,\bar{c}_j $, the action of the CUG is the operations $ U^{u_1}_{1}...U^{u_k}_{k} $. Any other basis state corresponds to the action of the CUG being the identity operator. The sum of the $ 2^{i+j} $ permutations yields the matrix of the $ C^{c_1,..., c_i}\bar{C}^{\bar{c}_1,...,\bar{c}_j}U^{u_1}_{1}...U^{u_k}_{k} $ gate. 
For example,
$$ C^{3}\bar{C}^{1}U^{2}  = P_0 \otimes I_2 \otimes P_0 + P_0 \otimes U \otimes P_1 + P_1 \otimes I_2 \otimes P_0 + P_1 \otimes I_2 \otimes P_1 , $$
as graphically shown in Figure \ref{fig:CCbarUDecompose}.

\subsubsection{Reduction to its irreducible form}

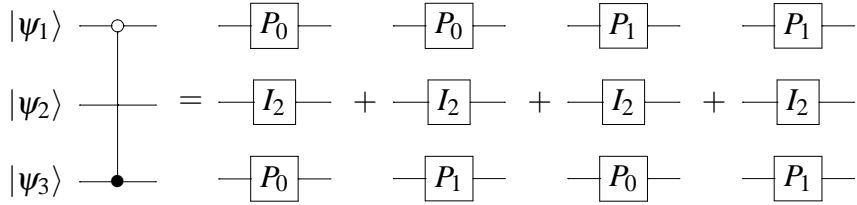
\begin{figure}[h]
\centering
\mbox
{
\Qcircuit @C=1.00em @R=2.30em 
{
	\lstick{\ket{\psi_1}} & \ctrlo{1} & \qw \\
	\lstick{\ket{\psi_2}} & \qw       & \qw \\
	\lstick{\ket{\psi_3}} & \ctrl{-1} & \qw
}
\qquad
\Qcircuit @C=1.00em @R=1.00em 
{
	           & \gate{P_0} & \qw \\
	\lstick{=} & \gate{I_2} & \qw \\
	           & \gate{P_0} & \qw
}
\qquad
\Qcircuit @C=1.00em @R=1.00em 
{
	           & \gate{P_0} & \qw \\
	\lstick{+} & \gate{I_2} & \qw \\
	           & \gate{P_1} & \qw
}
\qquad
\Qcircuit @C=1.00em @R=1.00em 
{
	           & \gate{P_1} & \qw \\
	\lstick{+} & \gate{I_2} & \qw \\
	           & \gate{P_0} & \qw
}
\qquad
\Qcircuit @C=1.00em @R=1.00em 
{
	           & \gate{P_1} & \qw \\
	\lstick{+} & \gate{I_2} & \qw \\
	           & \gate{P_1} & \qw
}
}
\caption{Decomposition of the $ I_8 $ gate.}
\label{fig:I8Decompose}
\end{figure}

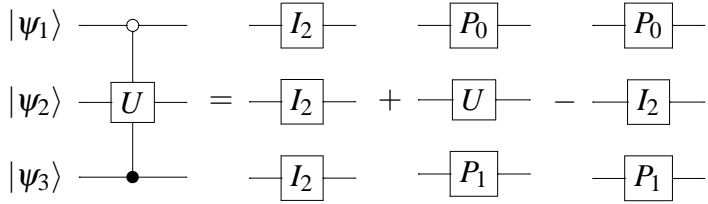
\begin{figure}[h]
\centering
\mbox
{
\Qcircuit @C=1.00em @R=1.60em 
{
	\lstick{\ket{\psi_1}} & \ctrlo{1} & \qw \\
	\lstick{\ket{\psi_2}} & \gate{U}  & \qw \\
	\lstick{\ket{\psi_3}} & \ctrl{-1} & \qw
}
\qquad
\Qcircuit @C=1.00em @R=1.00em 
{
	           & \gate{I_2} & \qw \\
	\lstick{=} & \gate{I_2} & \qw \\
	           & \gate{I_2} & \qw
}
\qquad
\Qcircuit @C=1.00em @R=1.00em 
{
	           & \gate{P_0} & \qw \\
	\lstick{+} & \gate{U}   & \qw \\
	           & \gate{P_1} & \qw
}
\qquad
\Qcircuit @C=1.00em @R=1.00em 
{
	           & \gate{P_0} & \qw \\
	\lstick{-} & \gate{I_2} & \qw \\
	           & \gate{P_1} & \qw
}
}
\caption{Optimized decomposition of the $ C^{3}\bar{C}^{1}U^{2} $ gate.}
\label{fig:CCbarUDecomposeOpt}
\end{figure}

For an arbitrary $ C^{c_1,..., c_i}\bar{C}^{\bar{c}_1,...,\bar{c}_j}U^{u_1}_{1}...U^{u_k}_{k} $ gate, a naive implementation of the previous section would require $ 2^{i+j} - 1 $ matrix additions to compute the matrix of the gate. However, this overhead can be reduced significantly by noting that only one permutation has the $ U^{u_1}_{1}...U^{u_k}_{k} $ operators being implemented, while the other $ 2^{i+j} - 1 $ possible permutations have identity operators substituted in for the $ U^{u_1}_{1}...U^{u_k}_{k} $ operators. As an example, consider the gate (essentially the identity matrix $I_8$) shown in Figure \ref{fig:I8Decompose}, which has $ 2^2 - 1 $ of the same permutations as in Figure \ref{fig:CCbarUDecompose}. Consequently, we can write the decomposition of the $ C^{3}\bar{C}^{1}U^{2} $ gate as the following 
\begin{equation}
C^{3}\bar{C}^{1}U^{2} = I_8 + P_0 \otimes U \otimes P_1 - P_0 \otimes I_2 \otimes P_1,
\end{equation}
which is graphically represented by Figure \ref{fig:CCbarUDecomposeOpt}.

\begin{figure}[h]
\centering
\mbox
{
\Qcircuit @C=1.00em @R=1.80em 
{
	\lstick{\ket{\psi_1}} & \ctrlo{1}      & \qw \\
	\lstick{\ket{\psi_2}} & \ctrl{1}       & \qw \\
	\lstick{\ket{\psi_3}} & \gate{U_1}     & \qw \\
	\lstick{\ket{\psi_4}} & \ctrlo{1} \qwx & \qw \\
	\lstick{\ket{\psi_5}} & \gate{U_2}     & \qw
}
\qquad
\Qcircuit @C=1.00em @R=1.00em 
{
	           & \gate{I_2} & \qw \\
	           & \gate{I_2} & \qw \\
	\lstick{=} & \gate{I_2} & \qw \\
	           & \gate{I_2} & \qw \\
	           & \gate{I_2} & \qw
}
\qquad
\Qcircuit @C=1.00em @R=1.00em 
{
	           & \gate{P_0} & \qw \\
	           & \gate{P_1} & \qw \\
	\lstick{+} & \gate{U_1} & \qw \\
	           & \gate{P_0} & \qw \\
	           & \gate{U_2} & \qw
}
\qquad
\Qcircuit @C=1.00em @R=1.00em 
{
	           & \gate{P_0} & \qw \\
	           & \gate{P_1} & \qw \\
	\lstick{-} & \gate{I_2} & \qw \\
	           & \gate{P_0} & \qw \\
	           & \gate{I_2} & \qw
}
}
\caption{Optimized decomposition of the $ C^{2}\bar{C}^{1,4}U^{3}_{1}U^{5}_{2} $ gate.}
\label{fig:CCbar2U12DecomposeOpt}
\end{figure}
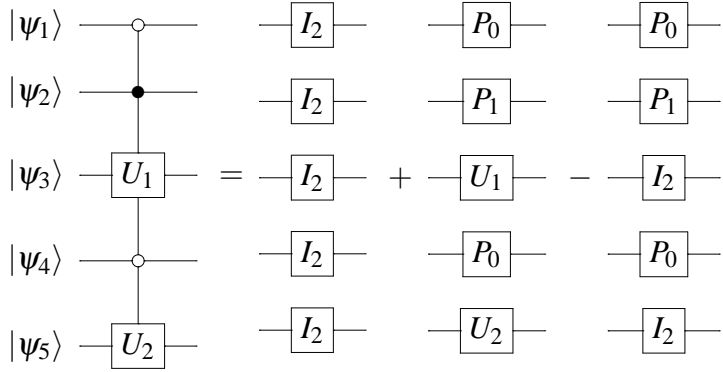

For the general case, the matrix of any arbitrary $ C^{c_1,..., c_i}\bar{C}^{\bar{c}_1,...,\bar{c}_j}U^{u_1}_{1}...U^{u_k}_{k} $ gate is simply the identity matrix (of appropriate dimensions), added together with the permutation that has the operations $ U^{u_1}_{1}...U^{u_k}_{k} $ implemented, subtracted with the same permutation with the $ U^{u_1}_{1}...U^{u_k}_{k} $ operators replaced with identity operators. In effect, the identity matrix is used to encapsulate $ 2^{i+j} - 1 $ permutations. Hence, for any arbitrary $ C^{c_1,..., c_i}\bar{C}^{\bar{c}_1,...,\bar{c}_j}U^{u_1}_{1}...U^{u_k}_{k} $ gate, computation of the gate matrix requires only two matrix additions, regardless of the number of controls or the gate dimensions. Note that the only instance in which this decomposition scheme is less efficient than the naive implementation is when only one $ C $ or $ \bar{C} $ conditional is involved. The optimized decomposition of a more complex example, the $ C^{2}\bar{C}^{1,4}U^{3}_{1}U^{5}_{2} $ gate, is given in Figure \ref{fig:CCbar2U12DecomposeOpt}.

\subsection{CUGs across qudits}

\subsubsection{Definitions and notation}

Denote the wavefuntion of the $i$-level qudit $j$ as $ \ket{\psi^{i}_{j}} $. Define a quantum circuit consisting of $n$ qudits $ \left\{ \ket{\psi^{\zeta_1}_{1}}, \ket{\psi^{\zeta_2}_{2}}, ... , \ket{\psi^{\zeta_n}_{n}} \right\} $ where $ \zeta_i $ represents the number of levels in the $i$th qudit and $ \zeta = \left\{ \zeta_1,\zeta_2,...,\zeta_n \right\} $. We call $\zeta$ the quantum circuit profile, which is the list of qudit levels, arranged according to the order of the qudits. For example, any CUG applied across qubits has $\zeta = \left\{2,2,...,2\right\}$, since qubits are 2-level qudits. Also define $ C^{c_i}_{s_i} $ as being conditional on the state $ \ket{s_i-1} $ of qudit $ c_i $, where $ 1 \leq s_i \leq \zeta_{c_i} $.

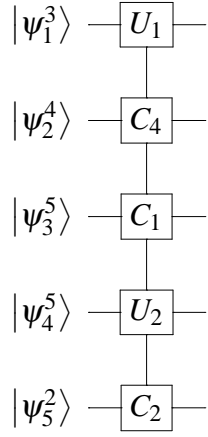
\begin{figure}[h]
\centering
\mbox
{
\Qcircuit @C=1.00em @R=1.60em 
{
	\lstick{\ket{\psi^{3}_1}} & \gate{U_1}      & \qw \\
	\lstick{\ket{\psi^{4}_2}} & \gate{C_4} \qwx & \qw \\
	\lstick{\ket{\psi^{5}_3}} & \gate{C_1} \qwx & \qw \\
	\lstick{\ket{\psi^{5}_4}} & \gate{U_2} \qwx & \qw \\
	\lstick{\ket{\psi^{2}_5}} & \gate{C_2} \qwx & \qw
}
}
\caption{The $ C^{2}_{4}C^{3}_{1}C^{5}_{2}U^{1}_{1}U^{4}_{2} $ gate, acting on 5 qudits of various levels. The quantum circuit profile is $ \zeta = \left\{3,4,5,5,2\right\} $.}
\label{fig:CUGQuditEg}
\end{figure}

A CUG with conditionals $ C^{c_1}_{s_1}...C^{c_i}_{s_i} $ implementing unitary operations $ U^{u_1}_{1}...U^{u_k}_{k} $, where $ u_1,...,u_k $ denotes the starting qudit of the corresponding $ U $ block, is represented by $ C^{c_1}_{s_1}...C^{c_i}_{s_i}U^{u_1}_{1}...U^{u_k}_{k} $. Figure \ref{fig:CUGQuditEg} shows an example of the $ C^{2}_{4}C^{3}_{1}C^{5}_{2}U^{1}_{1}U^{4}_{2} $ gate.

\subsubsection{Decomposition}

We can readily extend the concept of projection operators to qudits, by defining $ \left( P_{a,b} \right)_{i,j} = \delta_{ai}\delta_{aj} $ $ \forall $ $ 1 \leq i,j \leq b $ (where $ \delta_{ij} $ is the Kronecker delta) as the projection to the state $\ket{a-1}$ acting on a $b$-leveled qudit, with the restriction $ 1 \leq a \leq b $. Hence every $b$-leveled qudit has a set of $b$ projection operators defined, with the property $ \displaystyle\sum_{a=1}^{b} P_{a,b} = I_b $.

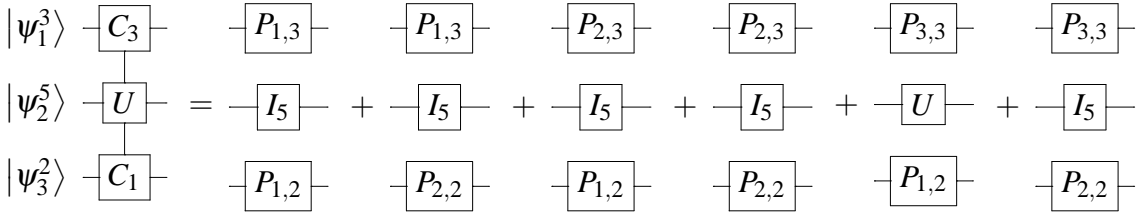
\begin{figure}[h]
\centering
\mbox
{
\Qcircuit @C=0.50em @R=1.00em 
{
	\lstick{\ket{\psi^{3}_1}} & \gate{C_3} & \qw \\
	\lstick{\ket{\psi^{5}_2}} & \gate{U} \qwx & \qw \\
	\lstick{\ket{\psi^{2}_3}} & \gate{C_1} \qwx & \qw
}
\qquad
\Qcircuit @C=0.50em @R=1.00em 
{
	           & \gate{P_{1,3}} & \qw \\
	\lstick{=} & \gate{I_5}     & \qw \\
	           & \gate{P_{1,2}} & \qw
}
\qquad
\Qcircuit @C=0.50em @R=1.00em 
{
	           & \gate{P_{1,3}} & \qw \\
	\lstick{+} & \gate{I_5}     & \qw \\
	           & \gate{P_{2,2}} & \qw
}
\qquad
\Qcircuit @C=0.50em @R=1.00em 
{
	           & \gate{P_{2,3}} & \qw \\
	\lstick{+} & \gate{I_5}     & \qw \\
	           & \gate{P_{1,2}} & \qw
}
\qquad
\Qcircuit @C=0.50em @R=1.00em 
{
	           & \gate{P_{2,3}} & \qw \\
	\lstick{+} & \gate{I_5}     & \qw \\
	           & \gate{P_{2,2}} & \qw
}
\qquad
\Qcircuit @C=0.50em @R=1.00em 
{
	           & \gate{P_{3,3}} & \qw \\
	\lstick{+} & \gate{U}   & \qw \\
	           & \gate{P_{1,2}} & \qw
}
\qquad
\Qcircuit @C=0.50em @R=1.00em 
{
	           & \gate{P_{3,3}} & \qw \\
	\lstick{+} & \gate{I_5}     & \qw \\
	           & \gate{P_{2,2}} & \qw
}
}
\caption{Decomposition of the $ C^{1}_{3}C^{3}_{1}U^{2} $ gate.}
\label{fig:C13C31UDecompose}
\end{figure}

For a general $ C^{c_1}_{s_1}...C^{c_i}_{s_i}U^{u_1}_{1}...U^{u_k}_{k} $ gate, it is clear that by substituting each conditional with a (valid) projection operator, it would result in $ \zeta_P = \displaystyle\prod_{j=1}^{i} \zeta_{c_j} $ distinct permutations. However, since the unitary operations $ U^{u_1}_{1}...U^{u_k}_{k} $ are only carried out iff the control qudits $ c_1,...,c_i $ are in the states $ \ket{s_1-1},...,\ket{s_i-1} $ respectively, only the permutation described by $ C^{c_j}_{s_j} \rightarrow P_{s_j,\zeta_{c_j}} $ $ \forall $ $ j = 1,...,i $ exactly will have 
$ U^{u_1}_{1}...U^{u_k}_{k} $ implemented; any other permutation will have identity operators substituted in place of $ U^{u_1}_{1}...U^{u_k}_{k} $. The sum of all $ \zeta_P $ permutations yields the matrix of the $ C^{c_1}_{s_1}...C^{c_i}_{s_i}U^{u_1}_{1}...U^{u_k}_{k} $ gate. For example, 

\begin{eqnarray*}
C^{1}_{3}C^{3}_{1}U^{2} & = & P_{1,3} \otimes I_5 \otimes P_{1,2} + P_{1,3} \otimes I_5 \otimes P_{2,2} + P_{2,3} \otimes I_5 \otimes P_{1,2} + \\
&& P_{2,3} \otimes I_5 \otimes P_{2,2} + P_{3,3} \otimes U \otimes P_{1,2} + P_{3,3} \otimes I_5 \otimes P_{2,2}
\end{eqnarray*}

as graphically shown in Figure \ref{fig:C13C31UDecompose}.

\subsubsection{Reduction to its irreducible form}

\begin{figure}[h]
\centering
\mbox
{
\Qcircuit @C=0.50em @R=1.00em 
{
	\lstick{\ket{\psi^{3}_1}} & \gate{C_3} & \qw \\
	\lstick{\ket{\psi^{5}_2}} & \gate{U} \qwx & \qw \\
	\lstick{\ket{\psi^{2}_3}} & \gate{C_1} \qwx & \qw
}
\qquad
\Qcircuit @C=1.00em @R=1.00em 
{
	           & \gate{I_3} & \qw \\
	\lstick{=} & \gate{I_5} & \qw \\
	           & \gate{I_2} & \qw
}
\qquad
\Qcircuit @C=0.50em @R=1.00em 
{
	           & \gate{P_{3,3}} & \qw \\
	\lstick{+} & \gate{U}   & \qw \\
	           & \gate{P_{1,2}} & \qw
}
\qquad
\Qcircuit @C=0.50em @R=1.00em 
{
	           & \gate{P_{3,3}} & \qw \\
	\lstick{-} & \gate{I_5}     & \qw \\
	           & \gate{P_{1,2}} & \qw
}
}
\caption{Optimized decomposition of the $ C^{1}_{3}C^{3}_{1}U^{2} $ gate.}
\label{fig:C13C31UDecomposeOpt}
\end{figure}
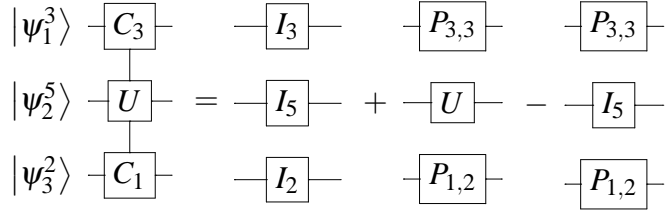

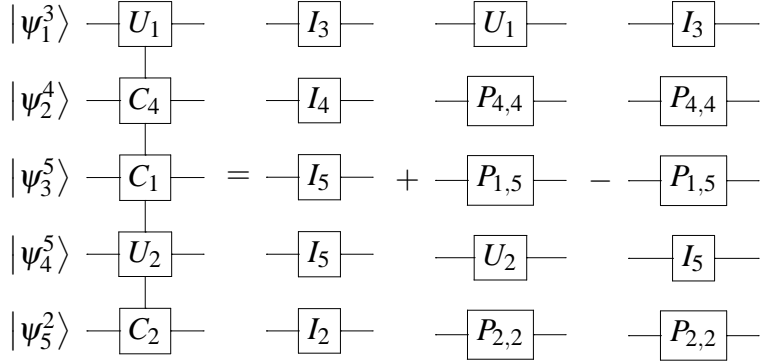
\begin{figure}[h]
\centering
\mbox
{
\Qcircuit @C=1.00em @R=1.00em 
{
	\lstick{\ket{\psi^{3}_1}} & \gate{U_1}      & \qw \\
	\lstick{\ket{\psi^{4}_2}} & \gate{C_4} \qwx & \qw \\
	\lstick{\ket{\psi^{5}_3}} & \gate{C_1} \qwx & \qw \\
	\lstick{\ket{\psi^{5}_4}} & \gate{U_2} \qwx & \qw \\
	\lstick{\ket{\psi^{2}_5}} & \gate{C_2} \qwx & \qw
}
\qquad
\Qcircuit @C=1.00em @R=1.00em 
{
	           & \gate{I_3} & \qw \\
	           & \gate{I_4} & \qw \\
	\lstick{=} & \gate{I_5} & \qw \\
	           & \gate{I_5} & \qw \\
	           & \gate{I_2} & \qw
}
\qquad
\Qcircuit @C=1.00em @R=0.95em 
{
	           & \gate{U_1} & \qw \\
	           & \gate{P_{4,4}} & \qw \\
	\lstick{+} & \gate{P_{1,5}} & \qw \\
	           & \gate{U_2} & \qw \\
	           & \gate{P_{2,2}} & \qw
}
\qquad
\Qcircuit @C=1.00em @R=0.95em 
{
	           & \gate{I_3}     & \qw \\
	           & \gate{P_{4,4}} & \qw \\
	\lstick{-} & \gate{P_{1,5}} & \qw \\
	           & \gate{I_5}     & \qw \\
	           & \gate{P_{2,2}} & \qw
}
}
\caption{Decomposition of the $ C^{2}_{4}C^{3}_{1}C^{5}_{2}U^{1}_{1}U^{4}_{2} $ gate.}
\label{fig:C24C31C52U11U42DecomposeOpt}
\end{figure}

As before, we can use the identity matrix (of appropriate dimensions) to encapsulate $ \zeta_P - 1 $ permutations of a $ C^{c_1}_{s_1}...C^{c_i}_{s_i}U^{u_1}_{1}...U^{u_k}_{k} $ gate, since only one of the permutations have the $ U^{u_1}_{1}...U^{u_k}_{k} $ operations implemented. The matrix of the CUG is thus the identity matrix (of appropriate dimensions) added together with the permutation described by $ C^{c_j}_{s_j} \rightarrow P_{s_j,\zeta_{c_j}} $ $ \forall $ $ j = 1,...,i $, minus the same permutation with identity matrices substituted in place of $ U^{u_1}_{1}...U^{u_k}_{k} $. The optimized decomposition of the $ C^{1}_{3}C^{3}_{1}U^{2} $ gate is given in Figure \ref{fig:C13C31UDecomposeOpt}. The optimized decomposition of a more complex example, the $ C^{2}_{4}C^{3}_{1}C^{5}_{2}U^{1}_{1}U^{4}_{2} $ gate, is given in Figure \ref{fig:C24C31C52U11U42DecomposeOpt}.

\section{Comparison with the \emph{QDENSITY} package}

The \emph{QDENSITY} package \cite{Diaz2006} provides many functions for the simulation of quantum circuits, two of which simulate the CNOT gate and the Toffoli gate. A more recent paper \cite{Tabakin2011} introduces \emph{QCWAVE} as an extension of the \emph{QDENSITY} package. \emph{QCWAVE} has the functions Op2 and Op3 that can be used to reproduce the action of $CNOT_n$ and $\mbox{Toffoli}_n$ gates on state vectors, but does not give the matrix of the gates itself.  We find it more straightforward and efficient to use the \emph{QDENSITY} functions to construct the matrix and then act on the state vector, and hence we perform the following comparison using the \emph{QDENSITY} package of version 4.0 (updated since \cite{Diaz2006}).

Here, we compare the CPU time taken to compute the matrices for the same gates, using the CNOT and Toffoli functions provided in the \emph{QDENSITY} package and the more general CUGate function provided in the \emph{CUGates.m} package. The \emph{QDENSITY} functions implements a decomposition using many more matrix additions and list manipulations in comparison with the scheme described in this paper.

\begin{figure}[h]
\begin{center}
\includegraphics[scale=0.32]{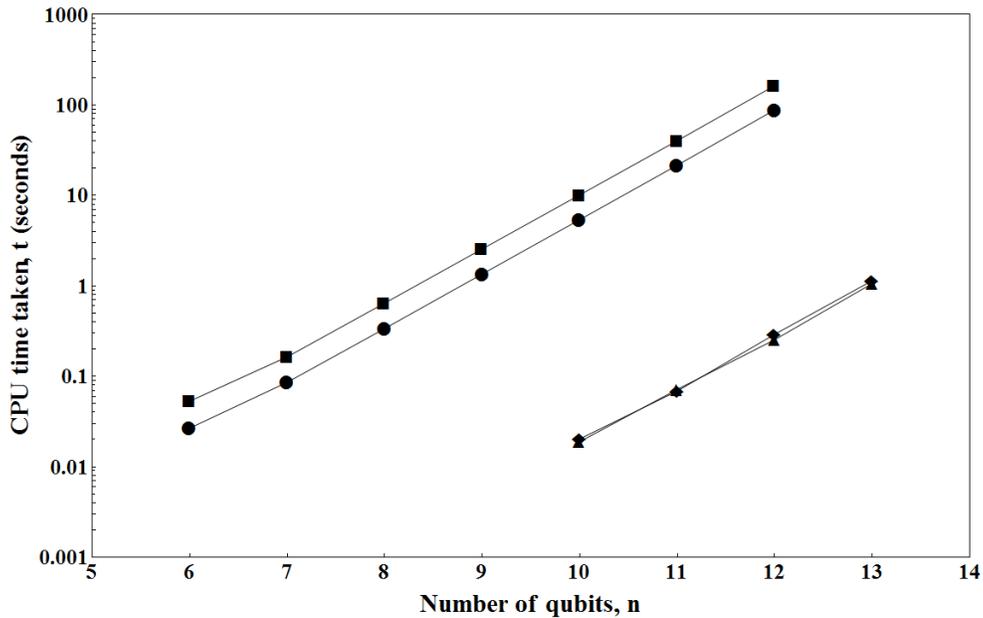}
\end{center}
\caption{CPU time taken. Circle/Square: Time taken using the CNOT/Toffoli function in the \emph{QDENSITY} 4.0 package. Diamond/Triangle: Time taken using the CUGate function in the \emph{CUGates.m} package with the sparse-matrix optimization to compute the CNOT/Toffoli gate.}
\label{fig:ComparisonLogPlot}
\end{figure}

We define the $CNOT_n$ gate as spanning $ n $ qubits with the $ C $ control located at the first qubit, and the NOT gate located at the $ n^{th} $ qubit. The $\mbox{Toffoli}_n$ gate is defined as spanning $ n $ qubits with the $ C $ controls at qubits 1 and 2, and the NOT gate located at the $ n^{th} $ qubit. Using these definitions, we are able to measure the CPU time taken to compute the matrix against $ n $, which is plotted in Figure \ref{fig:ComparisonLogPlot}.

As demonstrated in Figure \ref{fig:ComparisonLogPlot}, the CUGate function is significantly faster (by several orders of magnitude) than the CNOT and Toffoli functions provided in the QDENSITY package.  In the actual \emph{Mathematica} implementation of the CUGate function, we utilized sparse-matrix optimization to maximize calculation speed, which in this case, provides a speedup of about 1.8 compared to the CUGate function without the sparse-matrix optimization. It is also worth noting that while the Toffoli function takes almost twice as long as the CNOT function to compute its result, the CUGate function takes approximately the same length of time to compute the matrix of a CNOT and Toffoli gate for any particular $ n $, which is expected from the decomposition scheme described in this paper. In general, computation of the matrix of any two controlled unitary gates spanning the same number of qubits using the CUGate function takes the same length of time.

To perform this analysis, we have timed the use of the functions in \emph{Mathematica} using the Timing function, averaged over 10 trials. Computations were done on a laptop with an Intel Core i7-740QM processor with a speed of 1.73GHz.  Results for $ n < 10 $ 
using the CUGate function is omitted since the minimum granularity of the Timing function is more than the CPU time needed for the CUGate function. 

\section{Worked examples}

First load the CUGates.m package in \emph{Mathematica} using the following syntax:

\begin{fleqn}
\begin{align*}
\mbox{\footnotesize In[\theMathematicaIO]} := & \mbox{\textbf{ Needs[``CUGates`'']}}
\end{align*}
\end{fleqn}
\addtocounter{MathematicaIO}{1}

Brief descriptions of each function included in the CUGates.m package can be accessed using the `?' operator. For example, 

\begin{fleqn}
\begin{alignat*}{4}
\mbox{\footnotesize In[\theMathematicaIO]} := \mbox{ } & \mbox{\footnotesize\textbf{?CUGate}} \\
\mbox{\footnotesize Out[\theMathematicaIO]} := \mbox{ } & \mbox{\footnotesize CUGate[cpos,cbarpos,ubegin,umatrix]} \\
& \mbox{\footnotesize Returns the matrix of a CUG across qubits with C conditionals at cpos,} \\
& \mbox{\footnotesize $ \bar{C} $ conditionals at cbarpos, and unitary operators umatrix with the} \\
& \mbox{\footnotesize corresponding starting positions ubegin.}
\end{alignat*}
\end{fleqn}
\addtocounter{MathematicaIO}{1}

\begin{fleqn}
\begin{alignat*}{4}
\mbox{\footnotesize In[\theMathematicaIO]} := \mbox{ } & \mbox{\footnotesize\textbf{?CUGateG}} \\
\mbox{\footnotesize Out[\theMathematicaIO]} := \mbox{ } & \mbox{\footnotesize CUGateG[qcp,clist,ubegin,umatrix]} \\
& \mbox{\footnotesize Returns the matrix of a CUG across qudits with conditionals described by clist,} \\
& \mbox{\footnotesize and unitary operators umatrix with the corresponding starting positions ubegin.} \\
& \mbox{\footnotesize Note: clist is a list of \{Index of qudit in qcp,Conditional state\}}
\end{alignat*}
\end{fleqn}
\addtocounter{MathematicaIO}{1}

The qubit-specific subroutines are: BasisStateVector, CUGate, EqualSuperposition, HadamardGate, ListStates, MeasureQubits, MeasureSingleQubit, NOTGate, PHASEGate, SWAPGate and SWAPQubits.

The general qudit subroutines are: BasisStateVectorG, CUGateG, EqualSuperpositionG, ListStatesG, PHASEGateG, POp, QFTMinus, QFTPlus, RMinus, RPlus and SWAPQudits. The definitions for the functions QFTMinus and QFTPlus are similar to that of the QFT operator defined in \cite{Ermilov2009}.

\subsection{Shor's algorithm}

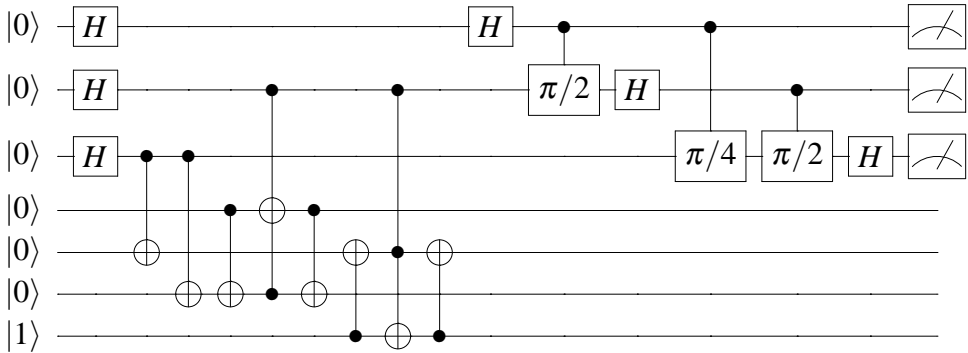
\begin{figure}[h]
\centering
\mbox
{
\Qcircuit @C=0.50em @R=0.50em
{
	\lstick{\ket{0}} & \gate{H} & \qw & \qw & \qw & \qw & \qw & \qw & \qw & \qw & \gate{H} & \ctrl{1} & \qw & \ctrl{2} & \qw & \qw & \meter \\
	\lstick{\ket{0}} & \gate{H} & \qw & \qw & \qw & \ctrl{2} & \qw & \qw & \ctrl{3} & \qw & \qw & \gate{\pi/2} & \gate{H} & \qw & \ctrl{1} & \qw & \meter \\
	\lstick{\ket{0}} & \gate{H} & \ctrl{2} & \ctrl{3} & \qw & \qw & \qw & \qw & \qw & \qw & \qw & \qw & \qw & \gate{\pi/4} & \gate{\pi/2} & \gate{H} & \meter \\
	\lstick{\ket{0}} & \qw & \qw & \qw & \ctrl{2} & \targ & \ctrl{2} & \qw & \qw & \qw & \qw & \qw & \qw & \qw & \qw & \qw & \qw \\
	\lstick{\ket{0}} & \qw & \targ & \qw & \qw & \qw & \qw & \targ & \ctrl{2} & \targ & \qw & \qw & \qw & \qw & \qw & \qw & \qw \\
	\lstick{\ket{0}} & \qw & \qw & \targ & \targ & \ctrl{-2} & \targ & \qw & \qw & \qw & \qw & \qw & \qw & \qw & \qw & \qw & \qw \\
	\lstick{\ket{1}} & \qw & \qw & \qw & \qw & \qw & \qw & \ctrl{-2} & \targ & \ctrl{-2} & \qw & \qw & \qw & \qw & \qw & \qw & \qw
}
}
\caption{Quantum circuit for Shor's algorithm, $ N = 15 $ and $ C = 7 $.}
\label{fig:ShorAlgorithmCircuit}
\end{figure}

Figure \ref{fig:ShorAlgorithmCircuit} shows the implementation of Shor's algorithm to factorize $ N = 15 $ for co-prime, $ C = 7 $ \cite{Vandersypen2001}.

Using \emph{Mathematica}, we first initialize the qubit states as follows:

\begin{fleqn}
\begin{alignat*}{4}
\mbox{\footnotesize In[\theMathematicaIO]} := \mbox{ } & \mbox{\footnotesize\textbf{InputVector }} = && \mbox{\footnotesize\textbf{ BasisStateVector[\{0,0,0,0,0,0,1\}];}} \\
& \mbox{\footnotesize\textbf{HTransform }} = && \mbox{ \footnotesize\textbf{KroneckerProduct[HadamardGate[],HadamardGate[],}} \\
&&& \mbox{\footnotesize\textbf{ HadamardGate[],IdentityMatrix[$2^{4}$]];}}
\end{alignat*}
\end{fleqn}
\addtocounter{MathematicaIO}{1}

Modular exponentiation is carried out on qubits 4 to 7 below:

\begin{fleqn}
\begin{alignat*}{4}
\mbox{\footnotesize In[\theMathematicaIO]} := \mbox{ } & \mbox{\footnotesize\textbf{ModA }} = && \mbox{\footnotesize\textbf{ KroneckerProduct[IdentityMatrix[$2^2$],}} \\
&&& \mbox{\footnotesize\textbf{ CUGate[\{3\},\{\},\{5\},\{NOTGate[]\}],IdentityMatrix[$2^2$]];}} \\
& \mbox{\footnotesize\textbf{ModB }} = && \mbox{\footnotesize\textbf{ KroneckerProduct[IdentityMatrix[$2^2$],}} \\
&&& \mbox{\footnotesize\textbf{ CUGate[\{3\},\{\},\{6\},\{NOTGate[]\}],IdentityMatrix[$2^1$]];}} \\
& \mbox{\footnotesize\textbf{ModC }} = && \mbox{\footnotesize\textbf{ KroneckerProduct[IdentityMatrix[$2^3$],}} \\
&&& \mbox{\footnotesize\textbf{ CUGate[\{4\},\{\},\{6\},\{NOTGate[]\}],IdentityMatrix[$2^1$]];}} \\
& \mbox{\footnotesize\textbf{ModD }} = && \mbox{\footnotesize\textbf{ KroneckerProduct[IdentityMatrix[$2^1$],}} \\
&&& \mbox{\footnotesize\textbf{ CUGate[\{2,6\},\{\},\{4\},\{NOTGate[]\}],IdentityMatrix[$2^1$]];}} \\
& \mbox{\footnotesize\textbf{ModE }} = && \mbox{\footnotesize\textbf{ ModC;}} \\
& \mbox{\footnotesize\textbf{ModF }} = && \mbox{\footnotesize\textbf{ KroneckerProduct[IdentityMatrix[$2^4$],}} \\
&&& \mbox{\footnotesize\textbf{ CUGate[\{7\},\{\},\{5\},\{NOTGate[]\}]];}} \\
& \mbox{\footnotesize\textbf{ModG }} = && \mbox{\footnotesize\textbf{ KroneckerProduct[IdentityMatrix[$2^1$],}} \\
&&& \mbox{\footnotesize\textbf{ CUGate[\{2,5\},\{\},\{7\},\{NOTGate[]\}]];}} \\
& \mbox{\footnotesize\textbf{ModH }} = && \mbox{\footnotesize\textbf{ ModF;}} 
\end{alignat*}
\end{fleqn}
\addtocounter{MathematicaIO}{1}

Next, the inverse QFT (Quantum Fourier Transform) is performed on the first three qubits:

\begin{fleqn}
\begin{alignat*}{4}
\mbox{\footnotesize In[\theMathematicaIO]} := \mbox{ } & \mbox{\footnotesize\textbf{QftA }} = \mbox{ } && \mbox{\footnotesize\textbf{KroneckerProduct[HadamardGate[],IdentityMatrix[$2^2$]];}} \\
& \mbox{\footnotesize\textbf{QftB }} = && \mbox{\footnotesize\textbf{KroneckerProduct[CUGate[\{1\},\{\},\{2\},\{PHASEGate[$\mathbf{\pi/2}$]\}],}} \\
&&& \mbox{\footnotesize\textbf{IdentityMatrix[$2^5$]];}} \\
& \mbox{\footnotesize\textbf{QftC }} = && \mbox{\footnotesize\textbf{KroneckerProduct[IdentityMatrix[$2^1$],HadamardGate[],}} \\
&&& \mbox{\footnotesize\textbf{IdentityMatrix[$2^5$]];}} \\
& \mbox{\footnotesize\textbf{QftD }} = && \mbox{\footnotesize\textbf{KroneckerProduct[CUGate[\{1\},\{\},\{3\},\{PHASEGate[$\mathbf{\pi/4}$]\}],}} \\
&&& \mbox{\footnotesize\textbf{IdentityMatrix[$2^4$]];}} \\
& \mbox{\footnotesize\textbf{QftE }} = && \mbox{\footnotesize\textbf{KroneckerProduct[IdentityMatrix[$2^1$],}} \\
&&& \mbox{\footnotesize\textbf{ CUGate[\{2\},\{\},\{3\},\{PHASEGate[$\mathbf{\pi/2}$]\}],IdentityMatrix[$2^4$]];}} \\
& \mbox{\footnotesize\textbf{QftF }} = && \mbox{\footnotesize\textbf{KroneckerProduct[IdentityMatrix[$2^2$],HadamardGate[],}} \\
&&& \mbox{\footnotesize\textbf{IdentityMatrix[$2^4$]];}}
\end{alignat*}
\end{fleqn}
\addtocounter{MathematicaIO}{1}

We then multiply the matrices together from right to left, apply it to an initial qubit states, and obtain the final state of the quantum register.

\begin{fleqn}
\begin{alignat*}{4}
\mbox{\footnotesize In[\theMathematicaIO]} := \mbox{ } & \mbox{\footnotesize\textbf{TMatrix }} = \mbox{\footnotesize\textbf{QftF.QftE.QftD.QftC.QftB.QftA.ModH.ModG.}} \\
& \settowidth{\fxargpush}{\mbox{\footnotesize\textbf{ TMatrix = }}} \mbox{\footnotesize\textbf{ \hspace{\fxargpush} ModF.ModE.ModD.ModC.ModB.ModA.HTransform;}} \\
& \mbox{\footnotesize\textbf{OutputVector }} = \mbox{\footnotesize\textbf{TMatrix.InputVector;}} \\
& \mbox{\footnotesize\textbf{ListStates[OutputVector];}} \\
\mbox{\footnotesize Out[\theMathematicaIO]} := \mbox{ } & \mbox{\footnotesize List of qubit states with a non-zero amplitude:} \\
& \mbox{\footnotesize $\left( \frac{1}{4} \right) \ket{0000001} + \left( \frac{1}{4} \right) \ket{0000100} + \left( \frac{1}{4} \right) \ket{0000111} + \left( \frac{1}{4} \right) \ket{0001101} +$} \\
& \mbox{\footnotesize $\left( \frac{1}{4} \right) \ket{0010001} + \left( \frac{1}{4} \right) \ket{0010100} + \left( - \frac{1}{4} \right) \ket{0010111} + \left( - \frac{1}{4} \right) \ket{0011101} +$} \\
& \mbox{\footnotesize $\left( \frac{1}{4} \right) \ket{0100001} + \left( - \frac{1}{4} \right) \ket{0100100} + \left( \frac{i}{4} \right) \ket{0100111} + \left( - \frac{i}{4} \right) \ket{0101101} +$} \\
& \mbox{\footnotesize $\left( \frac{1}{4} \right) \ket{0110001} + \left( - \frac{1}{4} \right) \ket{0110100} + \left( - \frac{i}{4} \right) \ket{0110111} + \left( \frac{i}{4} \right) \ket{0111101}$}
\end{alignat*}
\end{fleqn}
\addtocounter{MathematicaIO}{1}

The most important part of the result is the state measurement of qubits 1, 2 and 3, which constitute the output register. Upon measurement, qubit 1 is solely in the computational basis $ \ket{0} $, whereas qubits 2 and 3 are in a mixture of both computational bases, $ \ket{0} $ and $ \ket{1} $. Written in reverse order, we have a superposition of the combined states $ \ket{000} $, $ \ket{010} $, $ \ket{100} $, and $ \ket{110} $ for the three qubits in the output register, which has a periodicity of $ p = 2 $. According to Shor's algorithm, the factors are then given by the greatest common divisor $ \left( gcd \right) $ of $ C^{\frac{2^{n-1}}{p}} \pm 1 $ and $ N $, where $ n = 3 $ is the number of qubits in the output register. Therefore $ gcd( C^{\frac{2^{n-1}}{p}} \pm 1, N ) = gcd( 7^{\frac{2^{3-1}}{2}} \pm 1, 15 ) = gcd( 7^2 \pm 1, 15 ) = 3,5  $, which are indeed the factors of $ N = 15 $.

\subsection{Quantum random walks}

Here, we are concerned with the quantum circuit implementation of quantum walks on highly symmetrical graphs. There exists different software packages that can implement quantum random walks across graphs, e.g. the QWalk package implements a quantum walk across 1-dimensional and 2-dimensional lattices \cite{Marquezino2008} and the qwViz package visualize a quantum walks on arbitrarily complex graphs \cite{Berry2011}, as well as various quantum state based physical implementation schemes such as described in \cite{Kia2008, Kia2009}, without reference to a circuit implementation of the graph.  However, we consider circuit implementations here to illustrate the use of the CUGates package.

\subsubsection{16-length cycle}

\begin{figure}[h]
\centerline{
\Qcircuit @C=0.9em @R=0.2em @!R{
&&&&&&&  &&&&& \qw & \multigate{3}{\textrm{incr}} & \multigate{3}{\textrm{decr}} & \qw \\
&&&&&&&  &&&&& \qw & \ghost{\textrm{incr}} & \ghost{\textrm{decr}} & \qw \\
\mbox{\includegraphics[width=3.5cm]{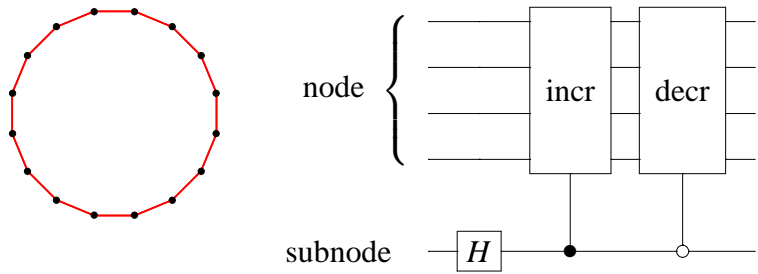}} &&&&&&& \ustick{\quad \; \textrm{node}} &&&&& \qw & \ghost{\textrm{incr}} & \ghost{\textrm{decr}} & \qw \\
&&&&&&& &&&&& \qw & \ghost{\textrm{incr}} & \ghost{\textrm{decr}} & \qw \\
&&&&&&& &&&&&&&&& \\
&&&&&&& \mbox{\quad \; \textrm{subnode}} &&&&& \gate{H} & \ctrl{-2} & \ctrlo{-2} & \qw
\gategroup{1}{11}{4}{11}{.3em}{\{}
}
}
\caption{\label{fig:cycle} Quantum circuit implementing a quantum walk along a 16-length cycle.}
\end{figure}

\begin{figure}[h]
\centerline{
\Qcircuit @C=1em @R=0.5em @!R{
&& \mbox{\quad \; Increment} &&& && && \mbox{\quad \; Decrement} &&& \\
& \targ \qw & \qw & \qw & \qw & \qw && & \targ \qw & \qw & \qw & \qw & \qw \\
&&&& \vdots & && &&&& \vdots & \\
& \ctrl{-2} & \targ \qw & \qw & \qw & \qw && & \ctrlo{-2} & \targ \qw & \qw & \qw & \qw \\
& \ctrl{-1} & \ctrl{-1} & \targ \qw & \qw & \qw && & \ctrlo{-1} & \ctrlo{-1} & \targ \qw & \qw & \qw \\
& \ctrl{-1} & \ctrl{-1} & \ctrl{-1} & \qswap \qw & \qw && & \ctrlo{-1} & \ctrlo{-1} & \ctrlo{-1} & \qswap \qw & \qw \\
}
}
\caption{\label{fig:IncrDecrImplement} Increment and decrement gates on $n$ qubits, producing cyclic permutations in the $2^n$ bit-string states.}
\end{figure}
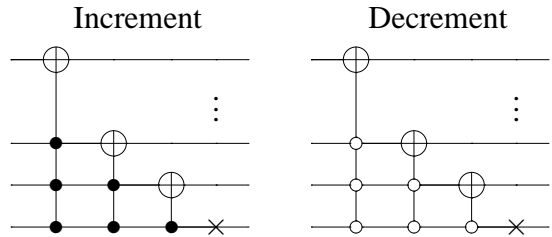

Consider the quantum circuit shown in Figure \ref{fig:cycle}, which implements a quantum walk on a 16-length cycle using the Increment/Decrement gates \cite{Douglas2009} shown in Figure \ref{fig:IncrDecrImplement}. First, we define the functions IncrementGate and DecrementGate in \emph{Mathematica} as below to calculate the matrix of the Increment/Decrement gate, given the number of qubits involved. 

\begin{fleqn}
\begin{alignat*}{4}
\mbox{\footnotesize In[\theMathematicaIO]} := & \settowidth{\fxargpush}{\mbox{\footnotesize\textbf{.}}} \mbox{\footnotesize\textbf{\hspace{\fxargpush}IncrementGate[NQubit\_Integer] :=}} \\
& \settowidth{\fxargpush}{\mbox{\footnotesize\textbf{.}}} \mbox{\footnotesize\textbf{\hspace{\fxargpush}(}} \\
& \settowidth{\fxargpush}{\mbox{\footnotesize\textbf{..}}} \mbox{\footnotesize\textbf{\hspace{\fxargpush}Module}} \\
& \settowidth{\fxargpush}{\mbox{\footnotesize\textbf{...}}} \mbox{\footnotesize\textbf{\hspace{\fxargpush}[\{ReturnMatrix,i,j\},}} \\
& \settowidth{\fxargpush}{\mbox{\footnotesize\textbf{......}}} \mbox{\footnotesize\textbf{\hspace{\fxargpush}ReturnMatrix = IdentityMatrix[$ 2^{NQubit} $];}} \\
& \settowidth{\fxargpush}{\mbox{\footnotesize\textbf{......}}} \mbox{\footnotesize\textbf{\hspace{\fxargpush}For[i = 1, i $<$ NQubit, $++$i,}} \\
& \settowidth{\fxargpush}{\mbox{\footnotesize\textbf{.........}}} \mbox{\footnotesize\textbf{\hspace{\fxargpush}ReturnMatrix = KroneckerProduct[IdentityMatrix[$2^{i-1}$],}} \\
& \settowidth{\fxargpush}{\mbox{\footnotesize\textbf{.........ReturnMatrix = }}} \mbox{\footnotesize\textbf{\hspace{\fxargpush}CUGate[Table[j,\{j,i+1,NQubit\}],\{\},\{i\},}} \\
& \settowidth{\fxargpush}{\mbox{\footnotesize\textbf{.........ReturnMatrix = }}} \mbox{\footnotesize\textbf{\hspace{\fxargpush}\{NOTGate[]\}].ReturnMatrix;}} \\
& \settowidth{\fxargpush}{\mbox{\footnotesize\textbf{......}}} \mbox{\footnotesize\textbf{\hspace{\fxargpush}];}} \\
& \settowidth{\fxargpush}{\mbox{\footnotesize\textbf{......}}} \mbox{\footnotesize\textbf{\hspace{\fxargpush}Return[KroneckerProduct[IdentityMatrix[$2^{NQubit-1}$],}} \\
& \settowidth{\fxargpush}{\mbox{\footnotesize\textbf{......Return.}}} \mbox{\footnotesize\textbf{\hspace{\fxargpush}NOTGate[]].ReturnMatrix];}} \\
& \settowidth{\fxargpush}{\mbox{\footnotesize\textbf{..}}} \mbox{\footnotesize\textbf{\hspace{\fxargpush}]}} \\
& \settowidth{\fxargpush}{\mbox{\footnotesize\textbf{.}}} \mbox{\footnotesize\textbf{\hspace{\fxargpush})}}
\end{alignat*}
\end{fleqn}
\addtocounter{MathematicaIO}{1}

\begin{fleqn}
\begin{align*}
\mbox{\footnotesize In[\theMathematicaIO]} := & \settowidth{\fxargpush}{\mbox{\footnotesize\textbf{.}}} \mbox{\footnotesize\textbf{\hspace{\fxargpush}DecrementGate[NQubit\_Integer] :=}} \\
& \settowidth{\fxargpush}{\mbox{\footnotesize\textbf{.}}} \mbox{\footnotesize\textbf{\hspace{\fxargpush}(}} \\
& \settowidth{\fxargpush}{\mbox{\footnotesize\textbf{..}}} \mbox{\footnotesize\textbf{\hspace{\fxargpush}Module}} \\
& \settowidth{\fxargpush}{\mbox{\footnotesize\textbf{...}}} \mbox{\footnotesize\textbf{\hspace{\fxargpush}[\{ReturnMatrix,i,j\},}} \\
& \settowidth{\fxargpush}{\mbox{\footnotesize\textbf{......}}} \mbox{\footnotesize\textbf{\hspace{\fxargpush}ReturnMatrix = IdentityMatrix[$ 2^{NQubit} $];}} \\
& \settowidth{\fxargpush}{\mbox{\footnotesize\textbf{......}}} \mbox{\footnotesize\textbf{\hspace{\fxargpush}For[i = 1, i $<$ NQubit, $++$i,}} \\
& \settowidth{\fxargpush}{\mbox{\footnotesize\textbf{.........}}} \mbox{\footnotesize\textbf{\hspace{\fxargpush}ReturnMatrix = KroneckerProduct[IdentityMatrix[$2^{i-1}$],}} \\
& \settowidth{\fxargpush}{\mbox{\footnotesize\textbf{.........ReturnMatrix = }}} \mbox{\footnotesize\textbf{\hspace{\fxargpush}CUGate[\{\},Table[j,\{j,i+1,NQubit\}],\{i\},}} \\
& \settowidth{\fxargpush}{\mbox{\footnotesize\textbf{.........ReturnMatrix = }}} \mbox{\footnotesize\textbf{\hspace{\fxargpush}\{NOTGate[]\}].ReturnMatrix;}} \\
& \settowidth{\fxargpush}{\mbox{\footnotesize\textbf{......}}} \mbox{\footnotesize\textbf{\hspace{\fxargpush}];}} \\
& \settowidth{\fxargpush}{\mbox{\footnotesize\textbf{......}}} \mbox{\footnotesize\textbf{\hspace{\fxargpush}Return[KroneckerProduct[IdentityMatrix[$2^{NQubit-1}$],}} \\
& \settowidth{\fxargpush}{\mbox{\footnotesize\textbf{......Return.}}} \mbox{\footnotesize\textbf{\hspace{\fxargpush}NOTGate[]].ReturnMatrix];}} \\
& \settowidth{\fxargpush}{\mbox{\footnotesize\textbf{..}}} \mbox{\footnotesize\textbf{\hspace{\fxargpush}]}} \\
& \settowidth{\fxargpush}{\mbox{\footnotesize\textbf{.}}} \mbox{\footnotesize\textbf{\hspace{\fxargpush})}}
\end{align*}
\end{fleqn}
\addtocounter{MathematicaIO}{1}

Using these definitions, we calculate the matrix of the circuit and apply it to the state vector signifying the initial vertex to be the 9th vertex (node representation of $ \ket{10000} $) with the subnode initially set to $ \ket{0} $.

\begin{fleqn}
\begin{alignat*}{4}
\mbox{\footnotesize In[\theMathematicaIO]} := \mbox{ } & \mbox{\footnotesize\textbf{InputVector = BasisStateVector[\{1,0,0,0,0\}];}} \\
& \mbox{\footnotesize\textbf{Coin = KroneckerProduct[IdentityMatrix[$2^4$],HadamardGate[]];}} \\
& \mbox{\footnotesize\textbf{T1 = CUGate[\{5\},\{\},\{1\},\{IncrementGate[4]\}];}} \\
& \mbox{\footnotesize\textbf{T2 = CUGate[\{\},\{5\},\{1\},\{DecrementGate[4]\}];}} \\
& \mbox{\footnotesize\textbf{TMatrix = T2.T1.Coin;}} \\
& \mbox{\footnotesize\textbf{OutputVector = TMatrix.InputVector;}} \\
& \mbox{\footnotesize\textbf{ListStates[OutputVector];}} \\
\mbox{\footnotesize Out[\theMathematicaIO]} := \mbox{ } & \mbox{\footnotesize List of qubit states with a non-zero amplitude:} \\
& \mbox{\footnotesize $ \left( \frac{1}{\sqrt{2}} \right) \ket{01100} + \left( \frac{1}{\sqrt{2}} \right) \ket{10011} $}
\end{alignat*}
\end{fleqn}
\addtocounter{MathematicaIO}{1}

From the output, we can see that the initial state $ \ket{10000} $ has been shifted to a superposition of states $ \ket{01100} $ and $ \ket{10011} $, which are the nodes adjacent to the initial state in a 16-length cycle. Further iterations will cause the quantum walk to propagate further along the cycle, with each state simultaneously moving to its adjacent states.

\subsubsection{Complete $ 3^3 $-graph with self-loops}

\begin{figure}[h]
\centerline
{
\Qcircuit @C=0.50em @R=0.30em @!R
{
	&&&&&&&&&&&&&&&&&&&&&&&&& \qw & \qw & \qw & \qw & \qw & \qw & \qw & \qswap \qwx[3] & \qw & \qw & \qw & \qw & \qw & \qw \\
	&&&&&&&&&&&&&&&&&&&&& \lstick{\textrm{Node}} &&&& \qw & \qw & \qw & \qw & \qw & \qw & \qw & \qw & \qw & \qswap \qwx[3] & \qw & \qw & \qw & \qw \\
	&&&&&&&&&&&&&&&&&&&&&&&&& \qw & \qw & \qw & \qw & \qw & \qw & \qw & \qw & \qw & \qw & \qw & \qswap \qwx[3] & \qw & \qw \\
	\mbox{\includegraphics[scale=0.20]{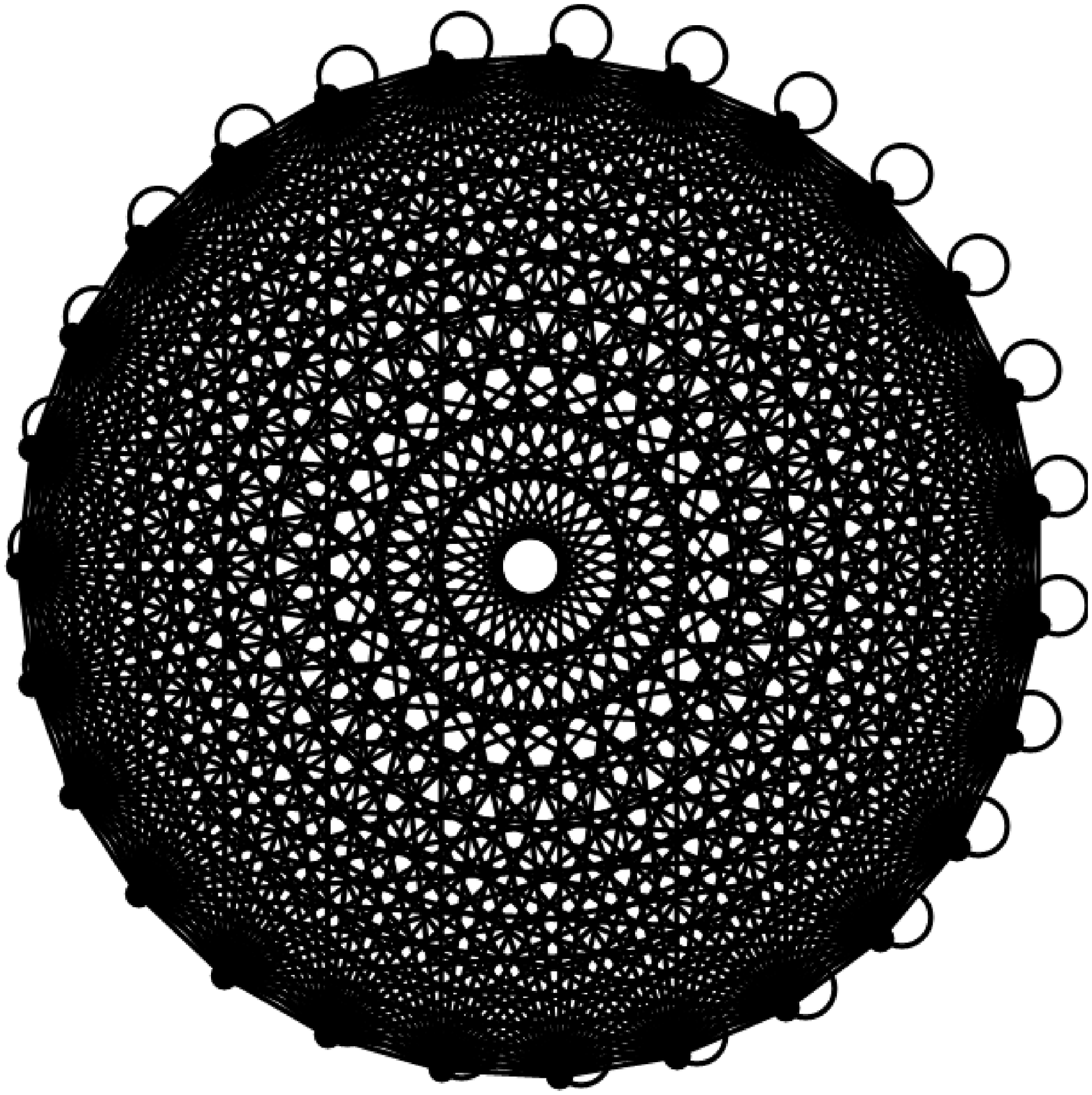}} &&&&&&&&&&&&&&&
	&&&&&&&&&& \qw & \gate{T_{+}} & \ctrlo{1} & \qw & \ctrlo{1} & \gate{T_{-}} & \qw & \qswap & \qw & \qw & \qw & \qw & \qw & \qw \\
	&&&&&&&&&&&&&&&&&&&&& \lstick{\textrm{Subnode}} &&&& \qw & \gate{T_{+}} & \ctrlo{1} & \qw & \ctrlo{1} & \gate{T_{-}} & \qw & \qw & \qw & \qswap & \qw & \qw & \qw & \qw \\
	&&&&&&&&&&&&&&&&&&&&&&&&& \qw & \gate{T_{+}} & \ctrlo{2} & \gate{\pi} & \ctrlo{2} & \gate{T_{-}} & \qw & \qw & \qw & \qw & \qw & \qswap & \qw & \qw \\
	&&&&&&&&&&&&&&&&&&&&&&&&&&&&&&&&&&&&&& \\
	&&&&&&&&&&&&&&&&&&&&&&&& \lstick{\ket{0}} & \qw & \qw & \targ & \ctrl{-2} & \targ & \qw & \qw & \qw & \qw & \qw & \qw & \qw & \qw & \qw 
	\gategroup{1}{25}{3}{25}{0.8em}{\{} 
	\gategroup{4}{25}{6}{25}{0.8em}{\{}
}
}
\caption{Quantum circuit implementing a quantum walk along a complete $ 3^3 $-graph with self-loops. The node and subnode are composed of 3-level qudits (i.e. qutrits).}
\label{fig:33CompleteGraph}
\end{figure}

As an example involving qudits in a quantum circuit, we analyze the quantum walk along the complete $ 3^n $-graph with self loops as discussed in \cite{Douglas2009}.  The complete $ 3^3 $-graph with self-loops can be constructed as in Figure \ref{fig:33CompleteGraph}.

Here, the operator $ T_{\pm} $ is defined as $ \left( T_{\pm} \right)_{a,b} = \frac{1}{\sqrt{3}} e^{\pm \frac{2\pi iab}{3}} $ $ \forall $ $ 1 \leq a,b \leq 3 $, and the quantum circuit profile is now $ \zeta = \{ 3,3,3,3,3,3,2 \} $. This can be implemented in \emph{Mathematica} as follows:

\begin{fleqn}
\begin{alignat*}{4}
\mbox{\footnotesize In[\theMathematicaIO]} := \mbox{ } & \mbox{\footnotesize\textbf{TMinus = QFTMinus[3];}} \\
& \mbox{\footnotesize\textbf{TPlus = QFTPlus[3];}} \\
& \mbox{\footnotesize\textbf{QCProfile = \{3,3,3,3,3,3,2\};}}
\end{alignat*}
\end{fleqn}
\addtocounter{MathematicaIO}{1}

The coin operator is calculated as follows:

\begin{fleqn}
\begin{alignat*}{4}
\mbox{\footnotesize In[\theMathematicaIO]} := \mbox{ } & \mbox{\footnotesize\textbf{C1 }} = && \mbox{\footnotesize\textbf{ SparseArray[KroneckerProduct[IdentityMatrix[$3^3$],}} \\
&&& \mbox{\footnotesize\textbf{ TPlus,TPlus,TPlus,IdentityMatrix[2]]];}} \\
& \mbox{\footnotesize\textbf{C2 }} = && \mbox{\footnotesize\textbf{ SparseArray[KroneckerProduct[IdentityMatrix[$3^3$],}} \\
&&& \mbox{\footnotesize\textbf{ CUGateG[QCProfile,\{\{4,1\},\{5,1\},\{6,1\}\},\{7\},\{NOTGate[]\}],}} \\
&&& \mbox{\footnotesize\textbf{ IdentityMatrix[2]]];}} \\
& \mbox{\footnotesize\textbf{C3 }} = && \mbox{\footnotesize\textbf{ SparseArray[KroneckerProduct[IdentityMatrix[$3^5$],}} \\
&&& \mbox{\footnotesize\textbf{ CUGateG[QCProfile,\{\{7,1\}\},\{6\},\{PHASEGateG[\{$\pi$,0,0\}]\}]]];}} \\
& \mbox{\footnotesize\textbf{C4 }} = && \mbox{\footnotesize\textbf{ C2;}} \\
& \mbox{\footnotesize\textbf{C5 }} = && \mbox{\footnotesize\textbf{ SparseArray[KroneckerProduct[IdentityMatrix[$3^3$],}} \\
&&& \mbox{\footnotesize\textbf{ TMinus,TMinus,TMinus,IdentityMatrix[2]]];}}
\end{alignat*}
\end{fleqn}
\addtocounter{MathematicaIO}{1}

The shifting operator can be implemented as such:

\begin{fleqn}
\begin{alignat*}{4}
\mbox{\footnotesize In[\theMathematicaIO]} := \mbox{ } & \mbox{\footnotesize\textbf{T1 }} = && \mbox{\footnotesize\textbf{ SparseArray[KroneckerProduct[SWAPQudits[QCProfile,{1,4}],}} \\
&&& \mbox{\footnotesize\textbf{ IdentityMatrix[$3^2*2$]]];}} \\
& \mbox{\footnotesize\textbf{T2 }} = && \mbox{\footnotesize\textbf{ SparseArray[KroneckerProduct[IdentityMatrix[3],}} \\
&&& \mbox{\footnotesize\textbf{ SWAPQudits[QCProfile,{2,5}], IdentityMatrix[$3*2$]]];}} \\
& \mbox{\footnotesize\textbf{T3 }} = && \mbox{\footnotesize\textbf{ SparseArray[KroneckerProduct[IdentityMatrix[$3^2$],}} \\
&&& \mbox{\footnotesize\textbf{ SWAPQudits[QCProfile,{3,6}], IdentityMatrix[2]]];}}
\end{alignat*}
\end{fleqn}
\addtocounter{MathematicaIO}{1}

Finally, we can calculate the matrix of the circuit, and apply it to the state vector signifying the initial vertex to be the 1st vertex (node representation of $ \ket{000} $), and obtain the result of a single iteration of the circuit.

\begin{fleqn}
\begin{alignat*}{4}
\mbox{\footnotesize In[\theMathematicaIO]} := \mbox{ } & \mbox{\footnotesize\textbf{InputVector = BasisStateVectorG[QCProfile,\{0,0,0,0,0,0,0\}]}} \\
& \mbox{\footnotesize\textbf{TMatrix = Normal[T3.T2.T1.C5.C4.C3.C2.C1];}} \\
& \mbox{\footnotesize\textbf{OutputVector = TMatrix.InputVector;}} \\
& \mbox{\footnotesize\textbf{ListStatesG[QCProfile,OutputVector];}} \\
\mbox{\footnotesize Out[\theMathematicaIO]} := \mbox{ } & \mbox{\footnotesize List of qudit states with a non-zero amplitude:} \\
& \mbox{\footnotesize $ \left( \frac{11}{27} \right) \ket{0000000} + \left( -\frac{16}{27} \right) \ket{0010000} + \left( -\frac{16}{27} \right) \ket{0020000} + \left( \frac{2}{27} \right) \ket{0100000} + $} \\
& \mbox{\footnotesize $ \left( \frac{2}{27} \right) \ket{0110000} + \left( \frac{2}{27} \right) \ket{0120000} + \left( \frac{2}{27} \right) \ket{0200000} + \left( \frac{2}{27} \right) \ket{0210000} + $} \\
& \mbox{\footnotesize $ \left( \frac{2}{27} \right) \ket{0220000} + \left( \frac{2}{27} \right) \ket{1000000} + \left( \frac{2}{27} \right) \ket{1010000} + \left( \frac{2}{27} \right) \ket{1020000} + $} \\
& \mbox{\footnotesize $ \left( \frac{2}{27} \right) \ket{1100000} + \left( \frac{2}{27} \right) \ket{1110000} + \left( \frac{2}{27} \right) \ket{1120000} + \left( \frac{2}{27} \right) \ket{1200000} + $} \\
& \mbox{\footnotesize $ \left( \frac{2}{27} \right) \ket{1210000} + \left( \frac{2}{27} \right) \ket{1220000} + \left( \frac{2}{27} \right) \ket{2000000} + \left( \frac{2}{27} \right) \ket{2010000} + $} \\
& \mbox{\footnotesize $ \left( \frac{2}{27} \right) \ket{2020000} + \left( \frac{2}{27} \right) \ket{2100000} + \left( \frac{2}{27} \right) \ket{2110000} + \left( \frac{2}{27} \right) \ket{2120000} + $} \\
& \mbox{\footnotesize $ \left( \frac{2}{27} \right) \ket{2200000} + \left( \frac{2}{27} \right) \ket{2210000} + \left( \frac{2}{27} \right) \ket{2220000} $}
\end{alignat*}
\end{fleqn}
\addtocounter{MathematicaIO}{1}

\subsubsection{$3^{rd}$ generation 3-Cayley tree}

\begin{figure}[h]
\begin{center}
\includegraphics[scale=0.60]{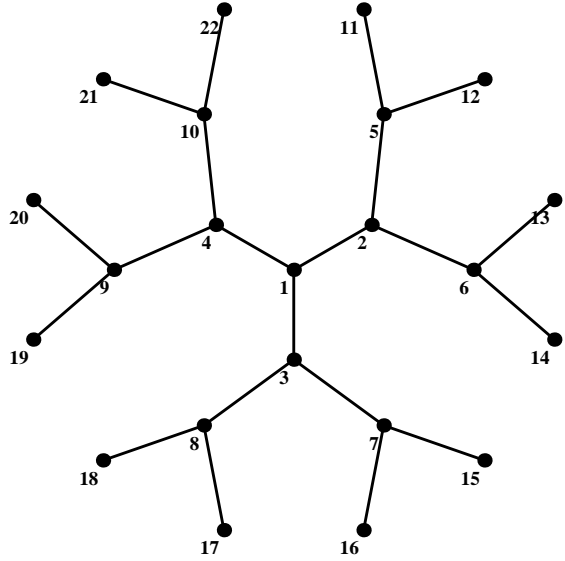}
\end{center}
\caption{$3^{rd}$ generation 3-Cayley tree.}
\label{fig:33Cayley}
\end{figure}

\begin{figure}[h]
\centerline{
\Qcircuit @C=0.23em @R=0.50em @!R
{
	&&&&&&&&&&&&
	\ctrlo{1} & \ctrl{1} & \qw & \ctrl{1} & \qw & \ctrlo{1} & \ctrlo{1} & \ctrl{1} & \multigate{1}{\textrm{incr}} & \qw & \multigate{1}{\textrm{decr}} & \qw & \ctrl{1} & \ctrlo{1} & \ctrlo{1} & \qw & \qw & \qw & \qw & \ctrlo{1} & \ctrlo{1} & \ctrlo{1} & \ctrlo{1} & \ctrlo{1} & \qw & \qw & \qw & \rstick{\ket{\psi^{2}_{1}}} \\
	& \ustick{\textrm{Level}} &&&&&&&&&&& 
	\ctrlo{1} & \ctrl{7} & \qw & \ctrl{7} & \qw & \ctrlo{7} & \ctrl{4} & \ctrl{4} & \ghost{incr} \qwx[2] & \qw & \ghost{decr} \qwx[2] & \qw & \ctrl{4} & \ctrlo{1} & \ctrlo{1} & \qw & \qw & \qw & \qw & \targ & \ctrlo{1} & \ctrlo{1} & \ctrlo{7} & \ctrl{4} & \qw & \qw & \qw & \rstick{\ket{\psi^{2}_{2}}} \\
	&&&&&&&&&&& \lstick{\mbox{Tree number}} &
	\gate{C_1} & \qw & \qw & \qw & \qw & \qw & \qw & \qw & \qw & \qw & \qw & \qw & \qw & \gate{\mathcal{R}_{+}} & \gate{\mathcal{R}_{-}} & \gate{C_1} & \qw & \gate{C_2} & \qw & \qw & \gate{\mathcal{R}_{-}} & \gate{\mathcal{R}_{+}} & \qw & \qw & \qw & \qw & \qw & \rstick{\ket{\psi^{3}_{3}}} \\
	&&&&&&&&&&&&
	\ctrlo{1} \qwx & \qw & \qw & \qw & \qw & \qw & \qw & \qw & \multigate{1}{R_L} & \qw & \multigate{1}{R_R} & \qw & \qw & \qw & \qw & \qw & \qw & \qw & \qw & \qw & \qw & \qw & \qw & \qw & \qw & \qw & \qw & \rstick{\ket{\psi^{2}_{4}}} \\
	& \ustick{\textrm{Node}} &&&&&&&&&&& 
	\ctrlo{1} & \qw & \qw & \qw & \ctrl{1} & \qw & \qw & \qw & \ghost{R_L} & \targ & \ghost{R_R} & \qw & \qw & \qw & \qw & \qw & \qw & \qw & \qw & \qw & \qw & \qw & \qw & \qw & \ctrl{1} & \qw & \qw & \rstick{\ket{\psi^{2}_{5}}} \\
	&&&&&&&&&&&&
	\multigate{1}{-G_3} & \qw & \multigate{1}{G_3} & \qw & \ctrl{1} & \qw & \ctrl{1} & \ctrlo{3} & \ctrlo{-1} & \qw & \ctrl{-1} & \targ & \ctrlo{3} & \ctrlo{-3} & \ctrl{-3} & \targ & \ctrlo{1} & \targ & \ctrlo{1} & \qw & \ctrlo{-3} & \ctrl{-3} & \qw & \ctrl{1} & \ctrl{1} & \qw & \qw & \rstick{\ket{\psi^{2}_{6}}} \\
	\ustick{\textrm{Subnode}} &&&&&&&&&&&& 
	\ghost{-G_3} & \qw & \ghost{G_3} & \qw & \targ & \qw & \ctrlo{2} & \qw & \qw & \ctrl{-2} & \qw & \qw & \qw & \ctrl{-1} & \ctrlo{-1} & \ctrlo{-4} & \targ & \ctrlo{-4} & \targ & \qw & \ctrl{-1} & \ctrlo{-1} & \qw & \ctrlo{2} & \targ & \qw & \qw & \rstick{\ket{\psi^{2}_{7}}} \\
	&&&&&&&&&&&&&&&&&&&&&&&&&&&&&&&&&&&&&& \\
	&&&&&&&&&&& \lstick{\ket{1^{2}}} & 
	\qw & \targ & \ctrl{-2} & \targ & \qw & \targ & \targ & \targ & \ctrl{-3} & \ctrl{-2} & \ctrl{-3} & \ctrl{-3} & \targ & \ctrlo{-2} & \ctrlo{-2} & \ctrlo{-2} & \qw & \ctrlo{-2} & \qw & \ctrlo{-7} & \ctrlo{-2} & \ctrlo{-2} & \targ & \targ & \qw & \qw & \qw & \rstick{\ket{1^{2}}} 
	\gategroup{1}{12}{2}{12}{0.8em}{\{} 
	\gategroup{4}{12}{5}{12}{0.8em}{\{}
	\gategroup{6}{12}{7}{12}{0.8em}{\{}
}
\hspace{5mm}
}
\caption{Quantum circuit implementing a quantum walk along a $3^{rd}$ generation 3-Cayley tree, with the central node marked. Any vertex is uniquely defined by a combination of the level, tree number, and node states.}
\label{fig:33CayleyQC}
\end{figure}
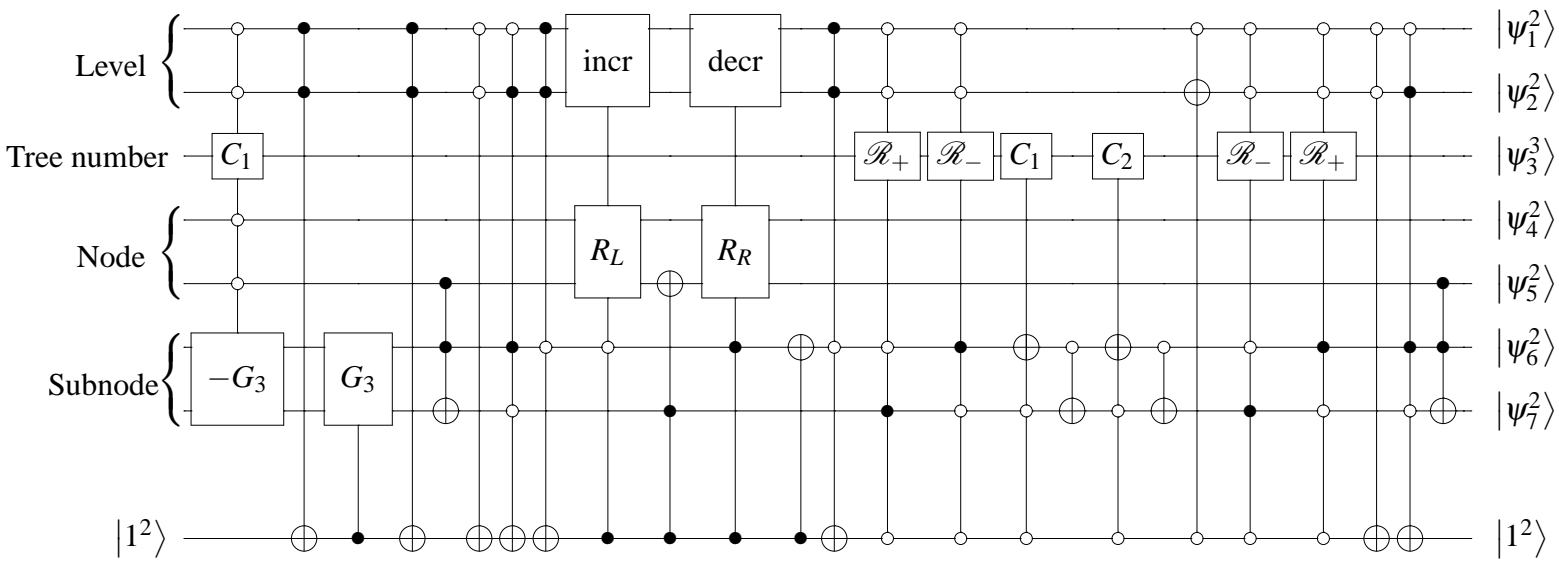

As a further example involving a mixture of qubits and qudits, we demonstrate how to implement a quantum walk on the $3^{rd}$ generation 3-Cayley tree (shown in Figure \ref{fig:33Cayley}) with the central node marked, by using its corresponding quantum circuit shown in Figure \ref{fig:33CayleyQC}.

\begin{figure}[h]
\centerline
{
\Qcircuit @C=0.9em @R=0.5em @!R{
	&& \mbox{\qquad \; $ \mathcal{R}_+ $ gate} &&& &&&&&&& && \mbox{\qquad \; $ \mathcal{R}_- $ gate} &&& \\
	\lstick{\ket{\psi^{\zeta_1}_{1}}} & \gate{\mathcal{R}_+} & \qw & \qw & \qw & \qw &&&&&&& \lstick{\ket{\psi^{\zeta_1}_{1}}} & \gate{\mathcal{R}_-} & \qw & \qw & \qw & \qw \\
	&&&& \vdots & &&&&&&& &&&& \vdots & \\
	\lstick{\ket{\psi^{\zeta_{n-2}}_{n-2}}} & \gate{C_{\zeta_{n-2}}} \qwx[-2] & \gate{\mathcal{R}_+} & \qw & \qw & \qw &&&&&&& \lstick{\ket{\psi^{\zeta_{n-2}}_{n-2}}} & \gate{C_{1}} \qwx[-2] & \gate{\mathcal{R}_-} & \qw & \qw & \qw \\
	\lstick{\ket{\psi^{\zeta_{n-1}}_{n-1}}} & \gate{C_{\zeta_{n-1}}} \qwx[-1] & \gate{C_{\zeta_{n-1}}} \qwx[-1] & \gate{\mathcal{R}_+} & \qw & \qw &&&&&&& \lstick{\ket{\psi^{\zeta_{n-1}}_{n-1}}} & \gate{C_{1}} \qwx[-1] & \gate{C_{1}} \qwx[-1] & \gate{\mathcal{R}_-} & \qw & \qw \\
	\lstick{\ket{\psi^{\zeta_{n}}_{n}}} & \gate{C_{\zeta_{n}}} \qwx[-1] & \gate{C_{\zeta_{n}}} \qwx[-1] & \gate{C_{\zeta_{n}}} \qwx[-1] & \gate{\mathcal{R}_+} & \qw &&&&&&& \lstick{\ket{\psi^{\zeta_{n}}_{n}}} & \gate{C_{1}} \qwx[-1] & \gate{C_{1}} \qwx[-1] & \gate{C_{1}} \qwx[-1] & \gate{\mathcal{R}_-} & \qw
}
}
\caption{$ \mathcal{R}_+ $ and $ \mathcal{R}_- $ gates on $ n $ qudits, with a quantum circuit profile of $ \zeta = \left\{ \zeta_1, \zeta_2, ..., \zeta_n \right\} $.}
\label{fig:RPRM}
\end{figure}
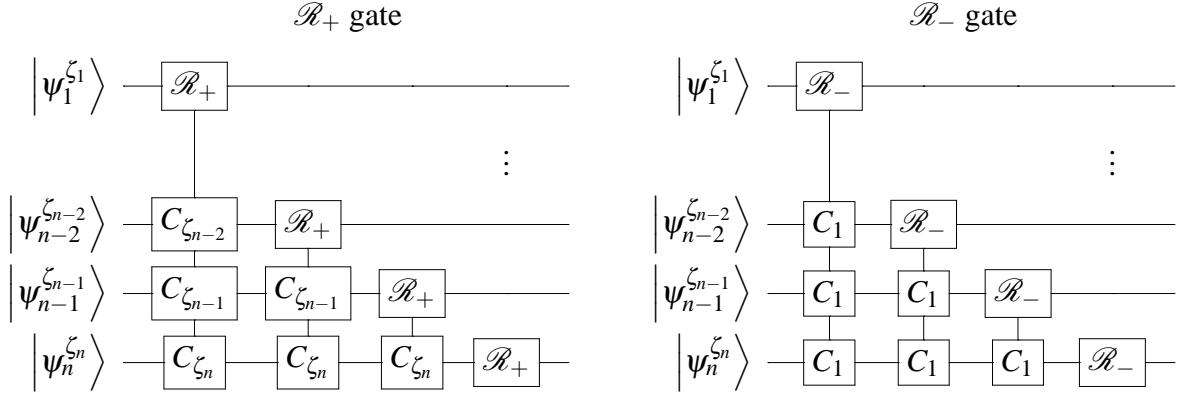

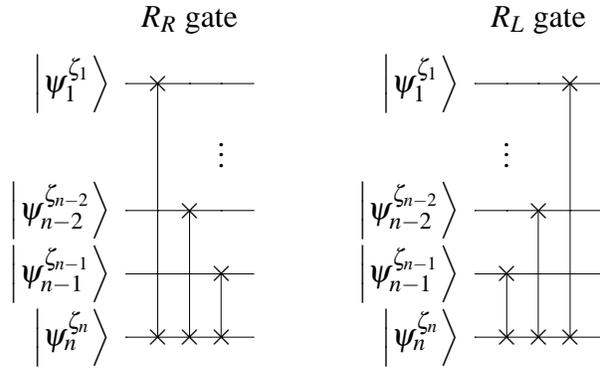
\begin{figure}[h]
\centerline
{
\Qcircuit @C=1.0em @R=2.0em @!{
	&& \mbox{$ R_R $ gate} &&& &&&&&&& & \mbox{$ R_L $ gate} &&&& \\
	\lstick{\ket{\psi^{\zeta_1}_{1}}} & \qswap & \qw & \qw & \qw &&&&&&& \lstick{\ket{\psi^{\zeta_1}_{1}}} & \qw & \qw & \qswap & \qw \\
	&&& \vdots &&& &&&&& & \vdots &&& \\
	\lstick{\ket{\psi^{\zeta_{n-2}}_{n-2}}} & \qw & \qswap  & \qw & \qw &&&&&&& \lstick{\ket{\psi^{\zeta_{n-2}}_{n-2}}} & \qw & \qswap & \qw & \qw \\
	\lstick{\ket{\psi^{\zeta_{n-1}}_{n-1}}} & \qw & \qw & \qswap & \qw &&&&&&& \lstick{\ket{\psi^{\zeta_{n-1}}_{n-1}}} & \qswap & \qw & \qw & \qw \\
	\lstick{\ket{\psi^{\zeta_{n}}_{n}}} & \qswap \qwx[-4] & \qswap \qwx[-2] & \qswap \qwx[-1] & \qw &&&&&&& \lstick{\ket{\psi^{\zeta_{n}}_{n}}} & \qswap \qwx[-1] & \qswap \qwx[-2] & \qswap \qwx[-4] & \qw
}
}
\caption{$ R_R $ and $ R_L $ gates on $ n $ qudits, with a quantum circuit profile of $ \zeta = \left\{ \zeta_1, \zeta_2, ..., \zeta_n \right\} $.}
\label{fig:RRRL}
\end{figure}

The $ G_n $ operator is defined as $ (G_n)_{i,j} = \frac{2}{n} - \delta_{ij} $ $ \forall $ $ 1 \leq i,j \leq n $. Here, the $ G_3 $ operator acts on only 3 of the 4 subnode states, so it does not mix with the state $ \ket{11} $. The $ \mathcal{R}_+ $ and $ \mathcal{R}_- $ gates are generalized increment and decrement gates respectively. For a $b$-leveled qudit, they are defined as $b$-by-$b$ matrices given as $ (\mathcal{R}_+)_{i,j} = \delta_{i(\mbox{mod b})+1,j} $ and $ (\mathcal{R}_-)_{i,j} = \delta_{i,j(\mbox{mod b})+1} $ respectively. A natural extension to multiple qudits is given in Figure \ref{fig:RPRM}. In general, the $ R_R $ and $ R_L $ operators, shown in Figure \ref{fig:RRRL}, correspond to a clockwise and anticlockwise rotation of qudits. However, in the context of Figure \ref{fig:33CayleyQC}, $ R_R $ and $ R_L $ are both single SWAP gates.

\begin{figure}[h]
\begin{center}
\includegraphics[scale=0.80]{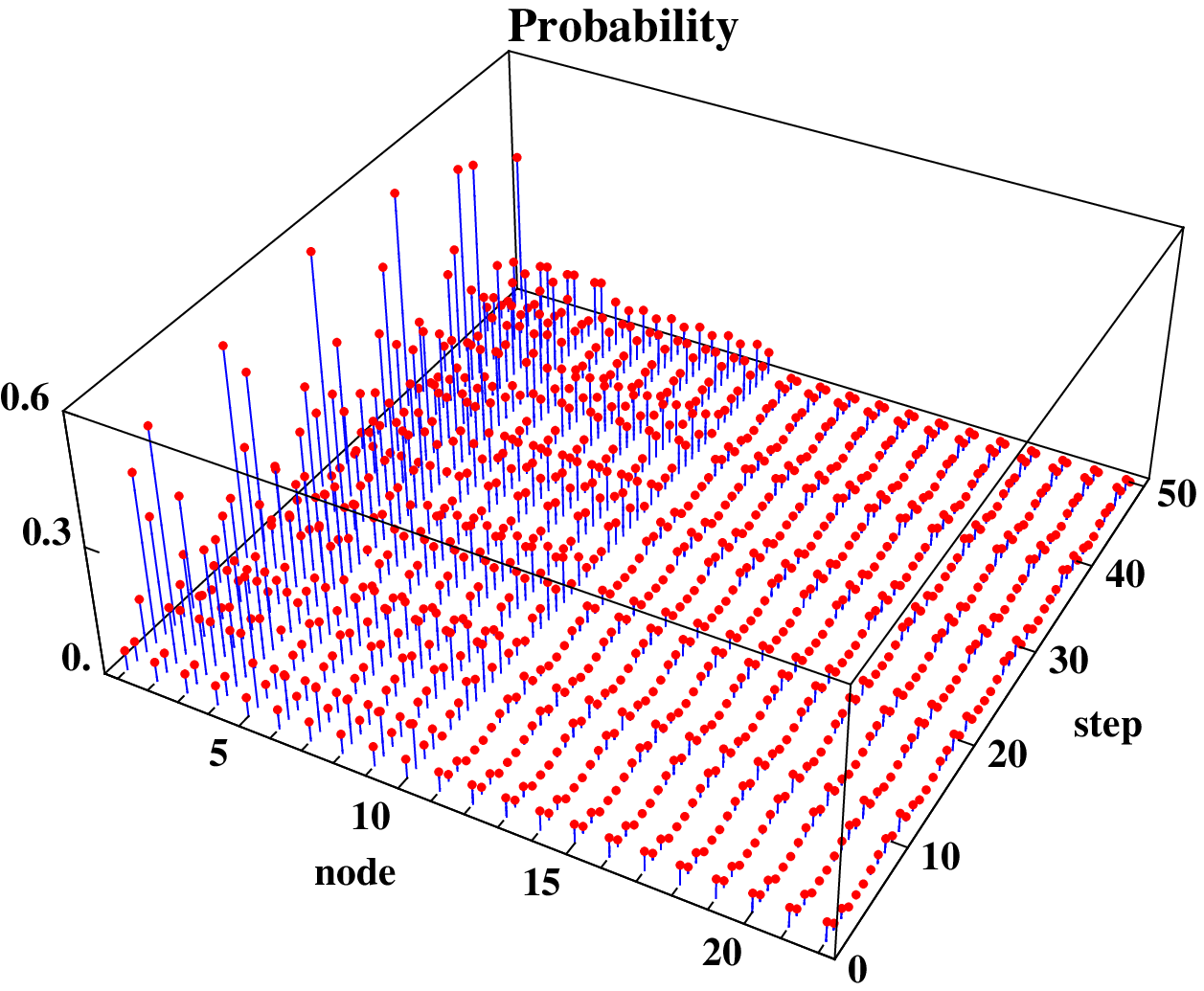}
\end{center}
\caption{Probability distribution along the $ 3^{rd} $ generation 3-Cayley tree against the number of walking steps. }
\label{fig:33CayleyResults}
\end{figure}

Given the length of the code needed to simulate the quantum circuit for a quantum walk along the 3-Cayley tree, we refer the reader to the \emph{Mathematica} notebook \emph{CUGates.nb}. The results of the quantum walk across 50 steps (starting in an equal superposition of vertex states, which is then subdivided according to the subnode states of the vertex) is shown in Figure \ref{fig:33CayleyResults}, where the centre marked node is distinguished by its much larger probability peak.

\section{Conclusions}

The \emph{Mathematica} notebook presented in this paper utilizes an irreducible form of matrix decomposition of a general controlled quantum gate with multiple conditionals and is highly efficient in simulating complex quantum circuits.  It provides a powerful tool to assist researchers analyze the performance of proposed quantum circuits. It has helped to identify several errors in the quantum circuits described in \cite{Douglas2009}, which was addressed and acknowledged in \cite{Douglas2009E}.  Another important application in which large and complex circuits need to be efficiently simulated is in the area of quantum error correction, in which generalized control unitary gates are used with both qubits and qudits \cite{Ionicioiu2009, Rabadi2009}.  This package will prove to be immensely helpful in the design of codification circuits in this area.  Implementation in \emph{Mathematica} allows the code to be used in a cohesive and interactive environment which is nevertheless computationally powerful. The interactive nature of this environment also makes this notebook suitable for teaching, where quantum algorithms and quantum gate operations can be studied in detail. 

\clearpage

\bibliographystyle{model1-num-names}

\bibliography{References}

\clearpage

\appendix
\appendixpage
\addappheadtotoc

\section{$ CU $ gate decomposition proof}

\begin{figure}[h]
\centering
\mbox
{
\Qcircuit @C=1.00em @R=2.00em
{
	\lstick{\ket{\psi_1}} & \gate{P_0} & \ctrl{1} & \qw \\
	\lstick{\ket{\psi_2}} & \qw        & \gate{U} & \qw 
}
\qquad
\Qcircuit @C=1.00em @R=1.00em
{
	           & \gate{P_0} & \qw &                           \\
	\lstick{=} &            &     & \rstick{=\mbox{ $ M_1 $}} \\
	           & \gate{I_2} & \qw & 
}
}
\caption{Application of $ P_0 $ to the $ CU $ gate.}
\label{fig:P1toCU}
\end{figure}
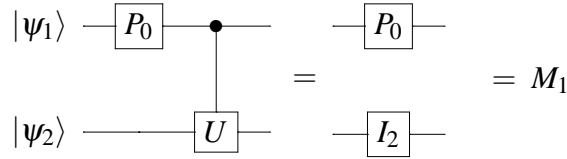

For any arbitary state, $ P_0 \left( a\ket{0} + b\ket{1} \right) \mapsto a\ket{0} $ and $ P_1 \left( a\ket{0} + b\ket{1} \right) \mapsto b\ket{1} $, i.e. the $ P_0 $ and $ P_1 $ operators projects arbitrary states onto the 
 $ \ket{0} $ and $ \ket{1} $ computational basis state respectively. Consider the quantum circuit in Figure \ref{fig:P1toCU}, where $ \ket{\psi_1} = a_1\ket{0} + b_1\ket{1} $ and $ \ket{\psi_2} = a_2\ket{0} + b_2\ket{1} $.

Since $ P_0 \left( \ket{\psi_1} \right) \mapsto a_1\ket{0} $, then  
$$ CU \left( P_0 \left( \ket{\psi_1} \right) \otimes \ket{\psi_2} \right) \mapsto a_1\ket{0} \otimes \ket{\psi_2} \equiv P_0 \left( \ket{\psi_1} \right) \otimes I_2 \left( \ket{\psi_2} \right) $$ 
i.e. the $ U $ gate is not applied to the second qubit because the control qubit is in the state $ a_1\ket{0} $ after the application of the $ P_0 $ operator, and thus the action of the $ CU $ gate is the identity operator. Hence, we can simplify the circuit, as shown in Figure \ref{fig:P1toCU}.

\begin{figure}[h]
\centering
\mbox
{
\Qcircuit @C=1.00em @R=2.00em
{
	\lstick{\ket{\psi_1}} & \gate{P_1} & \ctrl{1} & \qw \\
	\lstick{\ket{\psi_2}} & \qw        & \gate{U} & \qw 
}
\qquad
\Qcircuit @C=1.00em @R=1.00em
{
	           & \gate{P_1} & \qw &                            \\
	\lstick{=} &            &     & \rstick{=\mbox{ $ M_2 $}}  \\
	           & \gate{U}   & \qw &
}
}
\caption{Application of $ P_1 $ to the $ CU $ gate.}
\label{fig:P2toCU}
\end{figure}
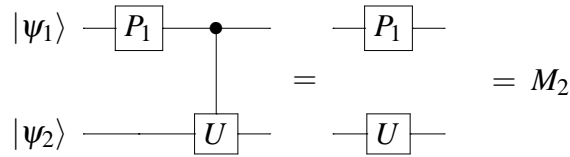

Similarly, if the $ P_1 $ operator is applied as in Figure \ref{fig:P2toCU}, then $ P_1 \left( \ket{\psi_1} \right) \mapsto b_1\ket{1} $ and thus 
$$ CU \left( P_1 \left( \ket{\psi_1} \right) \otimes \ket{\psi_2} \right) \mapsto b_1\ket{1} \otimes U \left( \ket{\psi_2} \right) \equiv P_1 \left( \ket{\psi_1} \right) \otimes U \left( \ket{\psi_2} \right), $$ 
because the control qubit is in the state $ b_1\ket{1} $ after the application of the $ P_1 $ operator, so the action of the $ CU $ gate is the $ U^2 $ operator. The equivalent circuit is also shown in Figure \ref{fig:P2toCU}.

Note that $ M_1 $ and $ M_2 $, as defined in Figures \ref{fig:P1toCU} and \ref{fig:P2toCU} respectively, are non-unitary. However the sum $ M_1 + M_2 = P_0 \otimes I_2 + P_1 \otimes U$ is unitary and also
\begin{eqnarray*}
M_1 + M_2 
& = & CU \left( P_0 \otimes I_2 \right) + CU \left( P_1 \otimes I_2 \right) \\
& = & CU \left( \left( P_0 + P_1 \right) \otimes I_2 \right) \\
& = & CU \left( I_2 \otimes I_2 \right) \\
& = & CU 
\end{eqnarray*}
Consequently, $CU = P_0 \otimes I_2 + P_1 \otimes U$.

\section{$ C^{1,3}U^{2} $ gate decomposition proof}

The decomposition can be derived by considering each of the possible permutations, which are defined as follows:
\begin{eqnarray*}
M_1 = C^{1,3}U^{2} \left( P_0 \otimes I_2 \otimes P_0 \right) = P_0 \otimes I_2 \otimes P_0  \\
M_2 = C^{1,3}U^{2} \left( P_0 \otimes I_2 \otimes P_1 \right) = P_0 \otimes I_2 \otimes P_1  \\
M_3 = C^{1,3}U^{2} \left( P_1 \otimes I_2 \otimes P_0 \right) = P_1 \otimes I_2 \otimes P_0  \\
M_4 = C^{1,3}U^{2} \left( P_1 \otimes I_2 \otimes P_1 \right) = P_1 \otimes U \otimes P_1 
\end{eqnarray*}
As before, we consider the permutation sum :
\begin{eqnarray*}
M_1 + M_2 + M_3 + M_4 
& = & C^{1,3}U^{2} \left( P_0 \otimes I_2 \otimes P_0 \right) + C^{1,3}U^{2} \left( P_0 \otimes I_2 \otimes P_1 \right) + \\
&   & C^{1,3}U^{2} \left( P_1 \otimes I_2 \otimes P_0 \right) + C^{1,3}U^{2} \left( P_1 \otimes I_2 \otimes P_1 \right) \\
& = & C^{1,3}U^{2} \left( P_0 \otimes I_2 \otimes \left( P_0 + P_1 \right) + {} \right. \\
&   & \left. P_1 \otimes I_2 \otimes \left( P_0 + P_1 \right) \right) \\
& = & C^{1,3}U^{2} \left( \left( P_0 + P_1 \right) \otimes I_2 \otimes I_2 \right) \\
& = & C^{1,3}U^{2} \left( I_2 \otimes I_2 \otimes I_2 \right) \\
& = & C^{1,3}U^{2} 
\end{eqnarray*}
Consequently, $ C^{1,3}U^{2} = P_0 \otimes I_2 \otimes P_0 + P_0 \otimes I_2 \otimes P_1 + P_1 \otimes I_2 \otimes P_0 + P_1 \otimes U \otimes P_1$.

\section{Arbitrary CUG decomposition}

For an arbitrary CUG across qubits with $ k $ conditionals, we have $ 2^k $ possible permutations when placing a $ P_0 $ or $ P_1 $ projection operator in front of each conditional. Each permutation then has a column that is described by the tensor product of the projection operators with identity matrices in the appropriate positions placed in front of the CUG. Proving that the sum of these permutations is equal to the gate itself is fairly trivial; it simply involves factoring together permutations that differ by a single conditional, using the identity $ P_0 + P_1 = I_2 $, and then doing so repeatedly until we end up with the original CUG. The simplification comes by considering the action of the projection operators on the state going into the CUG, and since the CUG implements the action iff the input state is in the basis state corresponding to the conditionals, we can easily work out which of the permutations has the action of the CUG implemented, while the rest do not.

Similarly, for an arbitrary CUG across qudits, we have a number of permutations corresponding to the qudit levels on which the conditionals are placed, and by using the identity $ \displaystyle\sum_{a=1}^{b} P_{a,b} = I_b $, we can readily prove that the sum of all permutations corresponds to the CUG itself, and can thus simplify the permutations as before. In both cases, we can simplify the decomposition considerably by using the identity matrix to represent the sum of all permutations with no action applied, then adding on the appropriate permutation with the action of the CUG and subtracting the same permutation without the action.

\end{document}